# Chemical Modification of Silicene[*]

Rong Wang (王蓉),[1] Mingsheng Xu (徐明生)[2] & Xiaodong Pi (皮孝东)[3][†]

[1]Key Laboratory of Interface Science and Engineering in Advanced Materials, Ministry of Education, Taiyuan University of Technology, Taiyuan 030024, China

[2]Department of Polymer Science and Engineering, Zhejiang University, Hangzhou 310027, China

[3]State Key Laboratory of Silicon Materials and School of Materials Science and Engineering, Zhejiang University, Hangzhou 310027, China

[†]E-mail: xdpi@zju.edu.cn

[*]Project supported by the 973 program (Grant No. 2013CB632101), the NSFC (Grant Nos. 61222404 & 61474097) and the Fundamental Research Funds for the Central Universities (2014XZZX003-09).

**ABSTRACT:** Silicene is a two-dimensional material, which is composed of a single layer of silicon atoms with $sp^2$-$sp^3$ mixed hybridization. The $sp^2$-$sp^3$ mixed hybridization renders silicene excellent reactive ability, facilitating the chemical modification of silicene. It has been demonstrated that chemical modification effectively enables the tuning of the properties of silicene. We now review all kind of chemical modification methods for silicene including hydrogenation, halogenation, organic surface modification, oxidation, doping and formation of two-dimensional hybrids. The effects of these chemical modification methods on the geometrical, electronic, optical and magnetic properties of silicene are discussed. The potential applications of chemically modified silicene in a variety of fields such as electronics, optoelectronics and magnetoelectronics are introduced. We finally envision future work on the chemical modification of silicene for the sake of further advancing the development of silicene.

## 1. Introduction

The rapid development of graphene has been stimulating great interest in producing all kinds of two-dimensional (2D) layered materials. Silicene, the two-dimensional honeycomb allotrope of silicon, has recently attracted intensive interest due to its unique properties [1, 2]. Charge carriers in silicene behave like massless Dirac Fermions, whose velocity near the Fermi level is estimated to be on the order of magnitude of $10^6$ m/s. This so-called Fermi velocity is comparable to that of graphene[3, 4]. The $sp^2$-$sp^3$ hybridization of Si atoms in silicene leads to a buckled structure of silicene, distinguishing silicene from the planar graphene. Such a distinct structure of silicene makes its properties different from those of graphene, creating new possibilities for manipulating the dispersion of electrons and enabling the tuning of the electronic and magnetic properties of silicene by external fields [6-8]. Silicene is predicted to possess strong spin-orbit coupling, which should result in detectable quantum spin Hall effect [5]. Moreover, silicene is essentially compatible with current Si-based technologies [9], giving rise to great promise in the electronic application of silicene[10-12]. Silicene field-effect transistors (FETs) exhibiting an ambipolar behavior with carrier mobilities around 100 cm$^2$/Vs at room temperature have already been demonstrated [11].

For the practical use of silicene, the opening and controlling of the bandgap of silicene are critical [12]. The reactive surface of silicene enables chemical modification to tune the band structure of silicene [13] by means of inorganic surface modification [14-38], organic surface modification [39-45], oxidation [46-53], doping [54-60] and formation of 2D hybrids [61-65]. Experimental investigation on the chemical modification of silicene is still at the preliminary stage [31, 40-47], while a great deal of theoretical investigation has been carried out to understand the chemical modification of silicene. The current review deals with the progress in the research on the chemical modification of silicene. Theoretical study on the effect of hydrogenation, halogenation, organic surface modification, oxidation, doping and formation of 2D hybrids on the electronic, optical and magnetic properties of silicene will be introduced in detail. Experimental work that mainly concern the oxidation and organic surface modification of silicene will be discussed. In the end, remarks and outlook are provided for the study of the chemical modification of silicene.

## 2. Inorganic surface modification

### 2.1 Hydrogenation

Hydrogenation is the process that hydrogen (H) atoms passivate silicene via the unpaired electrons of Si atoms [16]. Hydrogenation has been proved to be an effective chemical modification method to gain interesting properties for silicene. Thanks to the fine reversibility and excellent controllability of hydrogenation, a great deal of efforts have recently been devoted to tailoring the properties of silicene by hydrogenation [61-63]. In this section, we will deal with the formation, structures, electronic properties, optical properties and magnetic properties of hydrogenated silicene with full and half hydrogen passivation.

#### 2.1.1 Configurations of hydrogenated silicene

Fig. 1 shows that there are five possible atomic configurations (top-like, chair-like, boat-like, Z-line and A-line) for fully hydrogenated silicene [15, 17, 18]. Theoretical researchers have investigated the formation and stability of fully hydrogenated silicene with different configurations by comparing their formation and binding energies. The formation energy ($E_f$) and binding energy ($E_b$) of fully hydrogenated silicene were calculated by using $E_f = E_t - E_s - \frac{n_H}{2} E_{H_2}$ and $E_b = E_t - n_{Si} E_{Si} - n_H E_H$, respectively. $E_t$, $E_s$, $E_{H_2}$, $E_H$ and $E_{Si}$ were the total energy of fully hydrogenated silicene, silicene, hydrogen gas, a hydrogen atom and a silicon atom, respectively. $n_H$ and $n_{Si}$ were the number of H atoms and Si atoms, respectively [15, 17, 18]. It was found that the values of $E_f$ and $E_b$ were negative for all the structures of fully hydrogenated silicene, indicating the hydrogenation process was exothermic and the structures of fully hydrogenated silicene were stable [15, 17, 18]. The values of $E_f$ and $E_b$ for fully hydrogenated silicene with the chair-like configuration were the lowest. Therefore, it was concluded that the chair-like configuration was the most possible configuration for fully hydrogenated silicene in terms of formation and stability [15, 17].

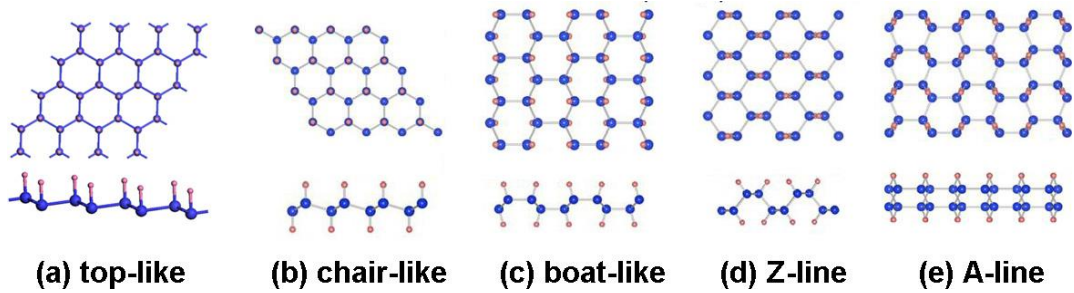

(a) top-like    (b) chair-like    (c) boat-like    (d) Z-line    (e) A-line

**Fig. 1** Optimized structures of fully hydrogenated silicene with (a) top-like, (b) chair-like, (c) boat-like, (d) Z-line and (e) A-line conforgurations. Silicon and hydrogen atoms are denoted by blue and pink balls, respectivelly[15, 17, 18].

Fig. 2 shows that half-hydrogenated silicene can be generated by the hydrogenation only at one side of silicene. Similar to what happens for the full hydrogenation of silicene, the zigzag, chair-like and boat-like configurations may form during the half hydrogenation of silicene (Fig. 2) [18]. Calculations on the $E_f$ and $E_b$ confirmed that the structures of half-hydrogenated silicene can all be formed and stable, as evidenced by the negative $E_f$ and $E_b$. The chair-like configuration is the most likely to form because of the lowest value of $E_f$ for half-hydrogenated silicene [18].

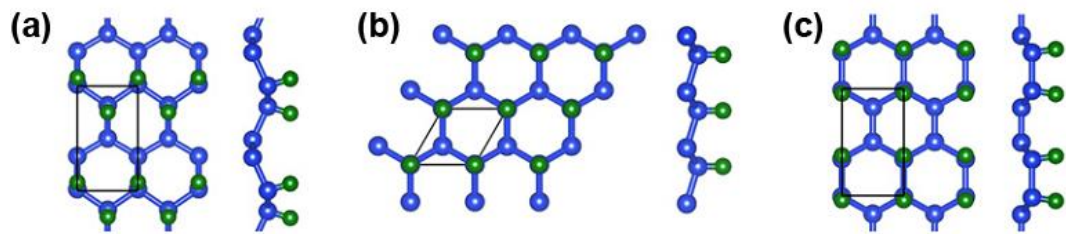

**Fig. 2** The optimized structures of half-hydrogenated silicene with (a) zigzag, (b) chair-like and (c) boat-like conforgurations. Silicon and hydrogen atoms are denoted by blue and green balls, respectivelly [18].

Although hydrogenated silicene with the chair-like configuration is the most likely to form, the reaction energy barrier for the hydrogenation of silicene is quite high (~1.8 eV) [19, 20]. It has been theoretically demonstrated that the energy barrier can be reduced by applying external tensile strain or electric field [19, 20]. Wu et al. have demonstrated that the energy barrier for the hydrogenation of silicene could be reduced by exerting biaxial strain or uniaxial tensile strain along the armchair direction of silicene [19]. Wu et al.'s results demonstrated that applying a perpendicular electric field could also facilitate the hydrogenation of silicene [20]. However, the external tensile strain or electric field changes the electronic structure of hydrogenated silicene because of the redistribution of atomic charges [19, 20]. It should be pointed out that tensile strain may also impact the electronic structure of silicene by changing the electron-phonon coupling [21, 22].

**2.1.2 Properties of fully hydrogenated silicene**

Since hydrogenated silicene is the most likely in the chair-like configuration, people usually focus on the properties of the chair-like hydrogenated silicene. The geometrical properties of fully hydrogenated silicene obtained with density functional theory (DFT) calculations are tabulated in Table 1. The buckling distance ($\Delta$), Si-Si bond length (<Si-Si>), Si-Si-Si bond angle ($\angle$(Si-Si-Si)) and Si-H bond length (<Si-H>) of fully hydrogenated silicene are 0.72-0.74 Å, 2.31-2.36 Å, 110° and 1.50-1.52 Å, respectively[15-18, 23, 24]. The buckling distance and Si-Si bond length of fully hydrogenated silicene are larger than those of silicene, while the Si-Si-Si bond angle is smaller than that of silicene. These structural changes indicate the hydrogenation-induced transition from mixed $sp^2$-$sp^3$ hybridization toward $sp^3$ hybridization in fully hydrogenated silicene for Si atoms.

Fig. 3 shows that fully hydrogenated silicene is an indirect-bandgap semiconductor with the valence-band maximum (VBM) and the conduction band minimum (CBM) appearing at the Γ and M points, respectively. The bandgap ($E_g$) obtained by DFT calculations with generalized gradient approximation (GGA) is 2.0 - 2.4 eV (Table 1) [15-18, 23, 24]. DFT calculations usually underestimate the bandgaps of semiconductors because the neglect of quasiparticle and excitonic effects [70]. Researchers have also performed hybrid Heyd–Scuseria–Ernzerhof (HSE) calculations as well as $G_0W_0$ (GW) many-body perturbation method for fully hydrogenated silicene. They have shown that the accurate bandgap of fully hydrogenated silicene were 3.5 - 4.0 eV (Table 1) [17, 18, 25].

**Table 1** Structural properties such as the buckling distance (Δ) of Si, the bond length of Si-Si, the bond length of Si-H, the bond angle of Si-Si-Si and bandgaps for the structures of silicene, fully hydrogenated silicene (FH-silicene) and half-hydrogenated silicene (HH-silicene).

|  | Δ (Å) | <Si-Si> (Å) | <Si-H> (Å) | ∠(Si-Si-Si) (°) | $E_g$ (eV) GGA | $E_g$ (eV) HSE06 |
|---|---|---|---|---|---|---|
| Silicene [3, 14, 69] | 0.45 | 2.28 |  | 116 | 0 | 0 |
| FH-silicene [16] | 0.72 | 2.319 | 1.502 |  | 2.00 [i] |  |
| FH-silicene [23] | 0.74 | 2.359 | 1.519 | 110.7 |  |  |
| FH-silicene [17] | 0.72 | 2.36 | 1.50 |  | 2.19 [i] |  |
| FH-silicene [18] |  | 2.36 | 1.50 |  | 2.36 [i] | 3.51 [i] |
| HH-silicene [18] |  | 2.35 | 1.51 |  | 0.84 [i] | 1.79 [d] |
| HH-silicene [32] | 0.61 | 2.34 | 1.51 |  | 0.95 [i] |  |

| | | | | |
|---|---|---|---|---|
| HH-silicene [33] | 0.69 | 2.35 | 1.52 | 0.93 [i] |

[i] indirect bandgap, [d] direct bandgap

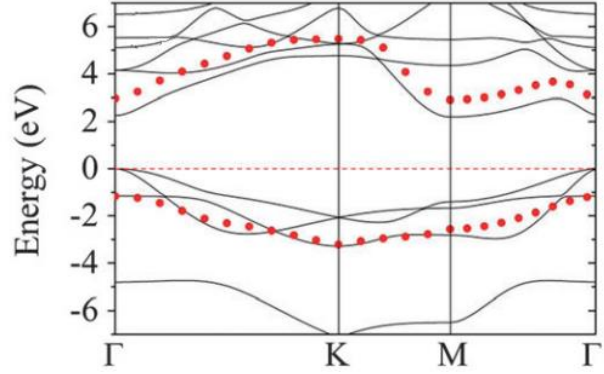

**Fig. 3** The band structure of fully hydrogenated silicene. Quasiparticle corrections (GW approximation) to the DFT energies are indicated as red dots. The Fermi level is at the energy of zero [25].

The optical absorption spectra of fully hydrogenated silicene calculated by DFT and GW approximation are shown in Fig. 4 (a) and (b). The absorption spectrum of fully hydrogenated silicene obtained with the DFT calculation shows an absorption onset at 2.2 eV, which is consistent with the bandgap of fully hydrogenated silicene calculated by using DFT. The absorption peak is located at 3.7 eV [26]. Wei et al. calculated the optical absorption spectra of fully hydrogenated silicene by using GW approximation with random phase approximation (GW+RPA) and Green's function perturbation theory (GW+BSE) [27]. The shape of the absorption spectrum obtained with GW+RPA was different from that obtained with DFT calculations because of the large self-energy correlations of electrons. The absorption onset obtained with GW+RPA located at ~ 4 eV was consistent with the bandgap. Since silicene is a two-dimensional material, the excitonic effect may introduce remarkable change in the optical properties of fully hydrogenated silicene. Therefore, Wei et al. calculated the optical absorption spectrum of fully hydrogenated silicene by using GW+BSE, which included the excitonic correlations of electrons and holes. Their results demonstrated that the excitonic effect significantly blueshifted the onset of the absorption spectrum of fully hydrogenated silicene. Strongly bound excitons below the onset of single-particle transition continuum appeared (Fig. 4 (b)). The inclusion of excitonic correlations between electrons and holes significantly redistributes the oscillator strength by introducing excitonic resonances. Since the excitonic effect significantly affects the optical properties of fully hydrogenated silicene, interactions of electrons and holes should be considered during the calculation of the optical absorption spectrum of fully hydrogenated silicene [27].

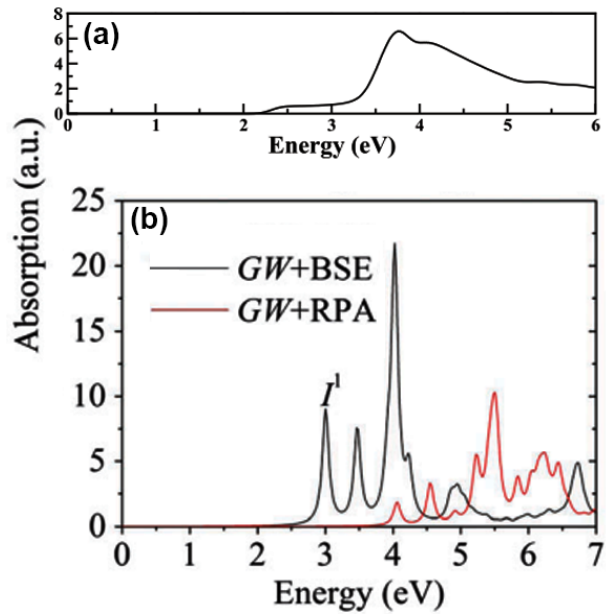

**Fig. 4** Optical absorption spectra of fully hydrogenated silicene calculated with (a) DFT calculations and (b) GW approximations with (GW + BSE) or without (GW + RPA) considering the excitonic effect [26, 27].

Based on its electronic and optical properties, fully hydrogenated silicene holds potential in optoelectronic applications in the ultraviolet (UV) region [27-29]. Although fully hydrogenated silicene is an indirect-bandgap semiconductor without linear band dispersion, a heterobilayer based on silicene and fully hydrogenated silicene turns to exhibit linear band dispersion character with a sizable band gap opening (44 − 61 meV) and high carrier mobility ($10^5$ cm$^2$ V$^{-1}$ s$^{-1}$) [30]. Therefore, fully hydrogenated silicene may be used as an inert substrate for silicene to make silicene transistors.

### 2.1.3 Observation of fully hydrogenated silicene

Qiu et al. investigated the hydrogenation of silicene by using scanning tunneling microscopy (STM) [31]. They synthesized silicene by evaporating silicon to a single-crystal Ag (111) substrate. They then exposed silicene to a high-purity $H_2$ gas at room temperature to generate hydrogenated silicene. It was found that the superstructure of silicene was 3×3 with the hexagonal arrangement of triangular structures around dark centers, as shown in Fig. 5 (a) [31]. The structural model of the 3×3 silicene indicated that each 3 × 3 unit cell was composed of two triangular half unit cells (Fig. 5 (b)). In each 3 × 3 unit cell there were 6 upper-buckled Si atoms indicated by red balls and 12 lower-buckled Si atoms indicated by yellow balls. The exposure of silicene to 900 L $H_2$ gas at room temperature resulted in hydrogenated silicene with invariable superstructure and rearranged buckling configuration. As shown in Fig. 5 (c), there were two different triangular half unit cells in each 3 × 3

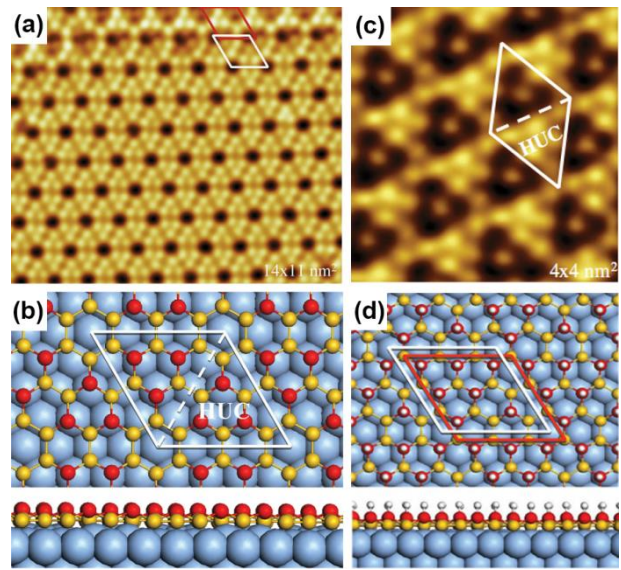

**Fig. 5** (a) The STM image and (b) the structural model of the clean silicene. (c) The STM image and (d) the structural model of the hydrogenated silicene [31].

unit cell of the hydrogenated silicene. As shown in the structural model of the hydrogenated silicene, one of the half unit cells had 6 bright spots while the other had only 1 bright spot in the middle (Fig. 5 (d)) [31]. The structure of hydrogenated silicene did not change when the $H_2$ dosage increased. This indicated that fully hydrogenated silicene was formed. Moreover, the dehydrogenation of fully hydrogenated silicene could occur by annealing at ~ 450 K. The superstructure and the buckling configuration of dehydrogenated silicene were the same as those of pristine silicene.

**2.1.4 Properties of half-hydrogenated silicene**

Half-hydrogenated silicene has the advantages of bandgap opening and novel magnetoelectric properties. Researchers usually focus on the properties of the chair-like half-hydrogenated silicene, because half-hydrogenated silicene is the most likely in the chair-like configuration. We now introduce the geometrical, electronic and magnetic properties of half-hydrogenated silicene. Table 1 shows that the buckling distance ($\Delta$), Si-Si bond length (<Si-Si>) and Si-H bond length (<Si-H>) of half-hydrogenated silicene with the chair-like configuration are 0.61 - 0.69 Å, 2.34 - 2.35 Å and 1.51 - 1.52 Å, respectively[18, 32, 33]. The buckling distance and Si-Si bond length of half-hydrogenated silicene are smaller than those of fully hydrogenated silicene, indicating that half-hydrogenated silicene is composed of mixed $sp^2$-$sp^3$ bonded Si atoms. After half hydrogenation, half Si atoms in silicene are passivated by H atoms via σ bonds, leaving electrons in unsaturated Si atoms localized and unpaired.

The band structure of half-hydrogenated silicene obtained with GGA calculations shows an indirect bandgap of 0.84 - 0.95 eV (Table 1) [18, 32, 33]. However, Fig. 6 shows that hybrid functional (HSE06) calculations indicate that half-hydrogenated silicene is a direct bandgap semiconductor with a bandgap of 1.79 eV[18]. As shown in the partial density of states (PDOS) and spatial spin-density distribution, states near the Fermi energy level are mainly derived from the *p* orbital of unsaturated Si atoms for half-hydrogenated silicene (Fig. 6 and Fig. 7) [18, 32].

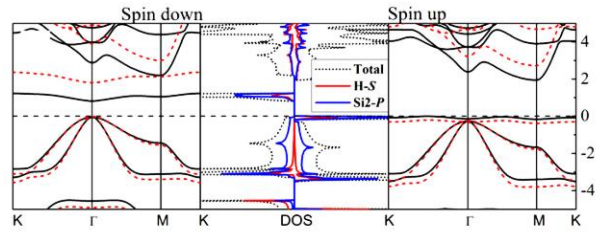

**Fig. 6** The band structure of half-hydrogenated silicene. The black solid and red dashed lines represent GGA and hybrid HSE06 results, respectively. The Fermi levels are shown by black dashed lines. The total and partial DOS of Si are given [18].

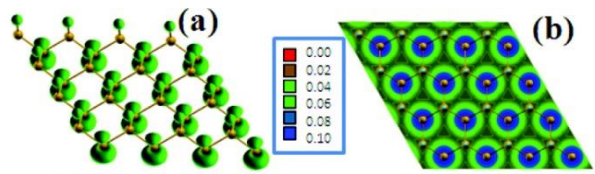

**Fig. 7** Spatial spin-density distribution of half-hydrogenated silicene. Side view and top view are presented in (a) and (b), respectively [32].

The spin-polarized DOS for the spin-up channel is significantly different from that for the spin-down channel in Fig. 6, indicating distinct magnetism in half hydrogenated silicene. This is because the $p_z$ electrons of unsaturated Si atoms are unpaired and localized, leading to a strong tendency to magnetize. However, the $p_z$ electrons of H-passivated Si atoms have no contribution to the magnetism (Fig. 7) [18, 32]. Pan et al. theoretically investigated the transport properties of FETs, in which half-hydrogenated silicene was used as the channel material (Fig. 8 (a)). The study showed that the on/off current ratio of a FET was as high as $10^6$, qualified for high-speed complementary MOS-like logic applications [33, 71] (Fig. 8 (b)). In addition, they predicted a spin-polarized current in the FET based on half-hydrogenated silicene. The spin-filter efficiency increased with the increase of gate voltage and reached 100% at the gate voltage of 1.9 V (Fig. 8 (c)) [33]. Tao et al. have recently fabricated FETs in which silicene was used as the channel material [11]. The on/off current ratios of their transistors were about one order of magnitude. They obtained the carrier mobility of ∼100 $cm^2V^{-1}s^{-1}$. Such a breakthrough in silicene-based FETs has laid a foundation for exploring all kinds of novel FETs based on chemically modified silicene including half-hydrogenated silicene.

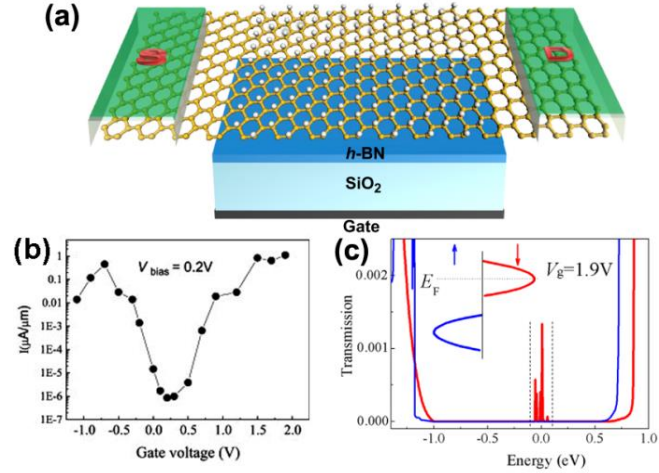

**Fig. 8** (a) Schematic of a FET based on half-hydrogenated silicene. The gate dielectric material is SiO$_2$. h-BN is the buffer layer. The electrodes are composed of semi-infinite silicene. The channel of half-hydrogenated silicene is 31.52 Å long. (b) Calculated transfer characteristic at the bias voltage of 0.2 V for the FET. (c) Spin-polarized transmission spectra with the gate voltage of 1.9 V. The bias voltage is fixed at 0.2 V. Red (blue) line stands for the spin down (up). The insets are the schematic of the Fermi energy level shift with the gate voltage [33].

## 2.2 Halogenation

Halogen (F, Cl, Br and I) atoms may passivate the surface of all kinds of Si materials such as bulk Si [72, 73], Si nanowires [74, 75] and Si quantum dots [76, 77]. As a novel Si material, silicene may be subjected to halogen passivation at the surface as well. Halogenation is the process that halogen atoms passivate silicene via the unpaired electrons of Si atoms at the surface. In the following we will introduce the structural, electronic, optical and magnetic properties of halogenated silicene with full and half passivation.

### 2.2.1 Configurations of halogenated silicene

Similar to what happens for the fully hydrogenation of silicene, the top, chair-like, boat-like, Z-line and A-line configurations may form during the full halogenation of silicene. The formation energies of fully halogenated silicene with these configurations are negative, indicating that they may be produced in the point of view of thermodynamics. The formation energy of fully halogenated silicene with the chair-like configuration is the lowest. Thus, it can be stated that the chair-like halogenated silicene is the most likely formed [35]. For fully halogenated silicene with a specific configuration, the formation energy increases in the order of F, Cl, Br and I. This means that halogenation less likely occurs as the atomic number of the halogen element increases [35].

As to the half halogenation of silicene, Wang et al. investigated the half-fluorination of silicene with zigzag, boat-like and chair-like configurations [36]. The optimized structures of half-fluorinated silicene with these configurations are shown in Fig. 9. They found that half-fluorinated silicene with these configurations were stable in the point of view of thermodynamics and dynamics. The $E_f$ of half-fluorinated silicene with the zigzag configuration is the lowest, indicating the zigzag configuration is the most likely to form.

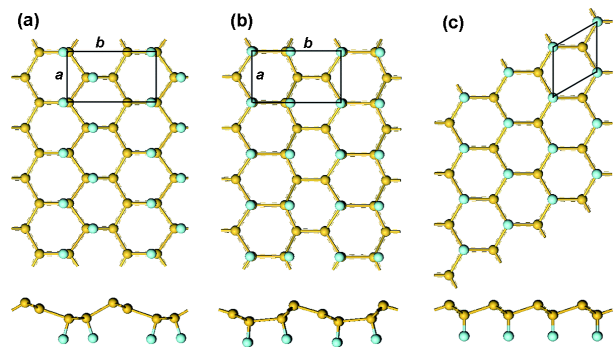

**Fig. 9** Optimized structures of half-fluorinated silicene with (a) zigzag, (b) boat-like, and (c) chair-like configurations with both top and side views. Silicon and fluorine atoms are denoted by yellow and blue balls, respectively [36].

**2.2.2 Properties of fully halogenated silicene**

We first discuss the geometrical, electronic and optical properties of fully halogenated silicene with the chair-like configuration. The geometrical properties of fully halogenated silicene are summarized in Table 2. The buckling distances of halogenated silicene are larger than that of silicene, implying that the hybridizations in halogenated silicene are closer to $sp^3$ hybridization than that in silicene [15, 27, 34-36]. The lengths of Si-Si bonds in halogenated silicene are larger than those in silicene and fully hydrogenated silicene. The bond length of Si-X (X = F, Cl, Br and I) increases with the increase of the atomic number of the halogen element, consistent with the change in the halogen-atom size [34, 35].

**Table 2.** Structural properties such as the buckling distance ($\Delta$) of Si, the bond length of Si-Si, the bond length of Si-X (X=F, Cl, Br and I) and bandgaps of fully halogenated silicene.

|  | METHOD | $\Delta$ (Å) | <Si-Si> (Å) | <Si-X> (Å) | $E_g$ (eV) |
|---|---|---|---|---|---|
| F-silicene | Ref [35] | 0.69 | 2.39 | 1.64 | 0.72 |
|  | Ref [34] | 0.69 | 2.40 | 1.63 | 1.47 |
|  | Ref [37] |  | 2.35 | 1.63 | $1.2^{[M06L]}, 1.1^{[VDW-DF]}, 1.6^{[HSE]}$ |
|  | Ref [15] | 0.70 | 2.39 | 1.63 | $0.82^d$ |
|  | Ref [28] | 0.70 | 2.40 | 1.62 | $0.49^{[LDA]}, 2.76^{[GW]}$ |
| Cl-silicene | Ref [35] | 0.72 | 2.39 | 2.08 | 1.28 |
|  | Ref [34] | 0.72 | 2.39 | 2.09 | 1.98 |
| Br-silicene | Ref [35] | 0.708 | 2.40 | 2.25 | 1.26 |

|  | Ref [34] | 0.705 | 2.41 | 2.23 | 1.95 |
| --- | --- | --- | --- | --- | --- |
| I-silicene | Ref [35] | 0.680 | 2.44 | 2.48 | 0.54 |
|  | Ref [34] | 0.677 | 2.45 | 2.47 | 1.19 |

All the fully halogenated silicene has a direct bandgap with both the VBM and CBM at the Γ point. The bandgap of fully halogenated silicene ranges from 0.54 to 2.76 eV, depending on halogen elements and calculation methods [15, 27, 34, 35]. As the passivating element changes from F to I, the bandgap of halogenated silicene firstly increases and then decreases. The bandgap of chlorinated silicene is the largest [34, 35]. Zhang et al. calculated the element-projected band structures of halogenated silicene [35]. Their results indicated that the CBM was insensitive to the halogen element, while the effect of the halogen element was appreciable for the VBM. Fig. 10 shows that the orbital contributions of the halogen element to the CBM are less than 20%. However, the orbital contributions of the halogen element to the VBM are < 20%, 20%, 40% and 80% for fluorination, chlorination, bromination and iodination, respectively [35]. Gao et al. have explained the orbital contributions by calculating the charge density distributions of the CBM and VBM states at the Γ point [34]. The CBM mainly originates from the antibonding state of the *p* orbitals of Si atoms, while the VBM originates from the *p* orbitals of Si atoms and halogen atoms. As the atomic number of the halogen element increases, the VBM is more significantly derived from the *p* orbitals of halogen atoms, indicating the formation of surface states upon halogenation. The interaction among the surface states is responsible for the dispersion of the CBM and VBM, and thus the opening of the bandgap [34].

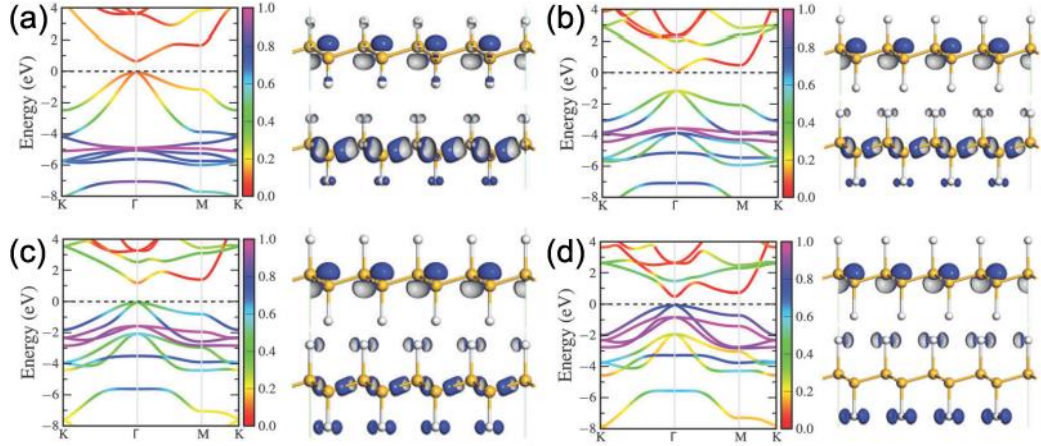

**Fig. 10** Band structures and charge density distributions of the CBM (right upper) and the VBM (right lower) states at the Γ point for silicene halogenated with (a) F, (b) Cl, (c) Br and (d) I [34, 35].

The effective carrier masses of fully halogenated silicene in certain directions are comparable to those of silicon [35]. Taking advantage of the tunable bandgap, relatively high carrier mobility and compatibility with silicon, fully halogenated silicene holds potential for electronic and optoelectronic applications.

**2.2.3 Properties of half-halogenated silicene**

Wang et al. have investigated the properties of half-fluorinated of silicene with the zigzag, boat-like and chair-like configurations [36]. They found that the electronic and magnetic properties of half-fluorinated silicene rigidly depended on the atomic configurations. Half-fluorinated silicene with zigzag and boat-like configurations are direct bandgap semiconductors with both VBM and CBM located at the Γ point. The GGA (HSE03)-calculated bandgaps are 0.19 (0.41) eV and 0.40 (0.82) eV for the structures of half-fluorinated silicene with the zigzag and boat-like configurations, respectively. Wang et al. have explained why half-fluorinated silicene with zigzag and boat-like configurations is semiconducting. As shown in Fig. 11 (a), the VBM of half-fluorinated silicene with the zigzag configuration results from both the $\sigma$ bonding states of Si-Si bonds and the $\pi$ bonding states of unsaturated Si-Si bonds, while its CBM mainly originates from the $p_z$ orbitals of unsaturated Si atoms and the anti-bonding states of Si-F bonds. For half-fluorinated silicene with the boat-like configuration, the VBM is only composed of the $\sigma$ bonding states of Si-Si bonds. The CBM is composed of the $p_z$ orbitals of unsaturated Si atoms and the anti-bonding states of Si-F bonds (Fig. 11 (b)). Thus, the weakening of the $sp^2$-$sp^3$ mixed coupling between Si atoms results in the semiconducting characteristics of half-fluorinated silicene with zigzag and boat-like configurations. The magnetic property of half-fluorinated silicene with the chair-like configuration is similar to that of half-hydrogenated silicene. The unpaired and localized $p_z$ electrons of unsaturated Si atoms results in spin polarization, which causes half-fluorinated silicene with the chair-like configuration to be antiferromagnetic [36].

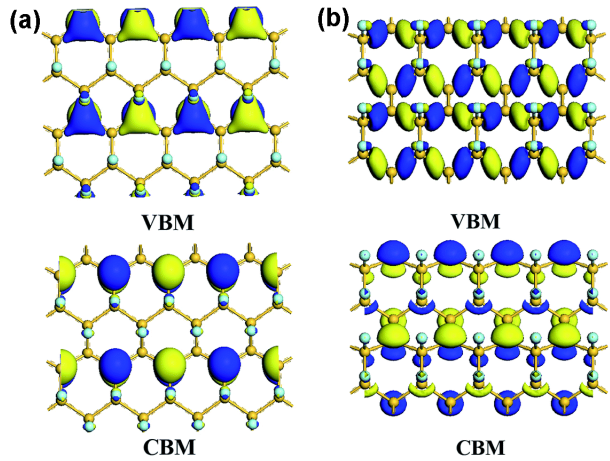

**Fig. 11** Partial charge densities of VBM and CBM for half-fluorinated silicene with (a) zigzag and (b) chair-like configurations [36].

Zheng et al. have studied the properties of half-brominated silicene with the chair-like configuration [38]. Fig. 12 shows its band structure, which indicates that it is semi-metallic. The energy bands of the spin-up channel close to the Fermi level indicate a semiconducting behavior, while those of the spin-down channel close to the

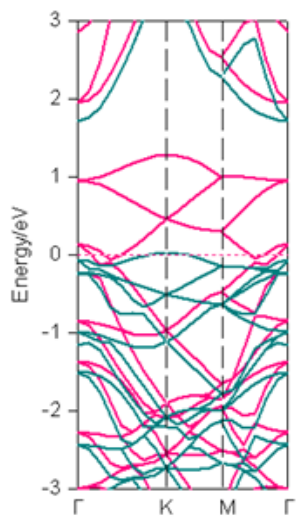

**Fig. 12** The band structure of half-brominated silicene. The spin-up and spin-down channels are denoted by green and pink lines, respectively [38].

Fermi level indicate a metallic behavior (Fig. 12). Therefore, 100% spin-polarized current can be obtained in half-brominated silicene, offering a feasible method to constructing spin devices based on silicene [38].

## 3. Organic surface modification

### 3.1 Experimental research

Nakano et al. have demonstrated the synthesis of silicene by the chemical exfoliation of calcium disilicide ($CaSi_2$) [78]. $CaSi_2$ has a hexagonal layered structure consisting of alternating Ca layers and corrugated silicon layers. The chemical exfoliation of $CaSi_2$ followed by HF rinsing leads to hydrogenated silicene [40]. Nakano et al. carried out hydrosilylation, aminization and phenylation for hydrogenated silicene to prevent it from oxidation, rendering the dispersibility of silicene in organic solvents and tailoring the optical properties of silicene [41-43]. Fig. 13 schematically shows Nakano et al.'s synthesis and organic surface modification of hydrogenated silicene [40-43, 78].

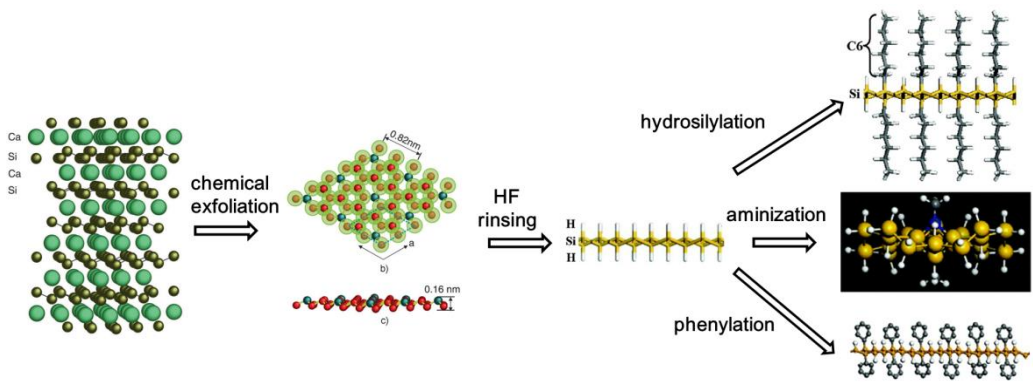

**Fig. 13** Schematic diagram of the synthesis and organic surface modification of hydrogenated silicene [40-43, 78].

Nakano et al. found that their hydrogenated silicene had stacked multi-layers. Hydrosilylated silicene was of single layer due to the steric hindrance of organic ligands. Moreover, hydrosilylation blueshifted the absorption onset of exfoliated silicene to 4.2 eV [41]. Aminizated silicene was found to be stacked to form multi-layers. Aminization introduced negligible effect on the absorption onset of hydrogenated silicene [42]. Phenylation separated hydrogenated silicene to be single-layered structure [43]. However, whether the change in the absorption spectrum was really induced by the hydrosilylation-induced surface chemistry was not clear because the number of the layers of hydrogenated silicene could not be well controlled.

### 3.2 Theoretical work

Theoretical study on the organic surface modification of silicene mainly deals with the formation and the properties of organically surface-modified silicene. Rubio-Pereda et al. have calculated the reaction energy barriers for the reactions between hydrogenated silicene and organic molecules such as acetylene, ethylene and styrene to evaluate the possibilities of the reactions [44]. The potential energy surface along the minimum energy path (MEP) that connects the intermediate state (IS), transition state (TS) and final state (FS) in a reaction has been considered. The IS, TS and FS correspond to the formation of a Si-C bond, H abstraction and the formation of a new C-H bond, respectively. Fig. 14 shows that during the reactions of acetylene, ethylene and styrene with hydrogenated silicene, the IS and FS are energetically more favorable than the TS. For acetylene, ethylene and styrene, the energy barriers of 0.20, 0.35 and 0.78 eV for H abstraction need to be circumvented to reach the FS of a reaction, respectively. The H-abstraction energy barriers are comparable or even smaller than those for the reactions of acetylene, ethylene and styrene with hydrogenated bulk Si (111) (0.54, 0.71 and 0.67 eV, respectively). The organic surface modification of hydrogenated bulk Si (111) has already been experimentally demonstrated. Therefore, organic surface modification is also possible for hydrogenated silicene [44].

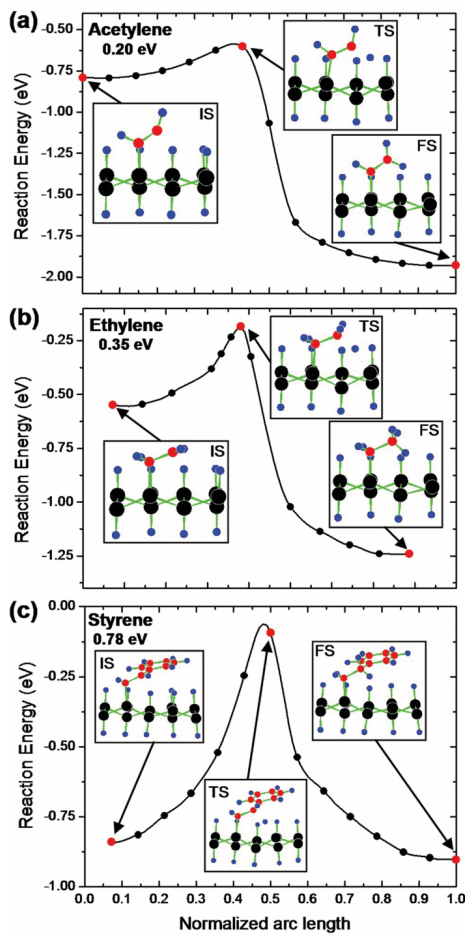

**Fig. 14** Potential energy surface along the MEP for the reaction of: (a) acetylene, (b) ethylene, and (c) styrene with the H-silicene. Silicon, carbon, and hydrogen atoms are represented by black, red, and blue balls, respectively [44].

Wang et al. have recently investigated the geometrical structures, band structures and optical absorption of hydrosilylated, alkoxylated, aminated and phenylated silicene by means of DFT calculations [39]. Propylene ($CH_2$=CH-$CH_3$), ethanol ($CH_3$-$CH_2$-OH), ethylamine (CH3-$CH_2$-$NH_2$) and phenylmagnesium bromide ($C_6H_5$-MgBr) were chosen for the simulation of hydrosilylation, alkoxylation, aminization and phenylation of silicene, respectively. Silicene was in fact initially passivated by hydrogen to facilitate all the organic surface modification. The surface coverage of organic ligands was consistent with the experimentally observed organic surface coverage of 33%. The optimized structures of hydrosilylated, phenylated, alkoxylated and aminated silicene are shown in Fig. 15. It was found that organic surface modification led to the increase of the buckling distance of hydrogenated silicene, while causing the angles of bonds in the honeycomb structure of silicene to decrease. Hydrosilylated and phenylated silicene were indirect bandgap semiconductors with the bandgap energy of 1.8 and 2.0 eV, respectively. Alkoxylated and aminated silicene were direct-bandgap semiconductors with the bandgap energies of ~ 1.7 eV. The authors also calculated the optical absorption spectra of hydrogenated silicene and organically surface modified silicene, as shown in Fig. 16. For hydrogenated silicene, the absorption onset was at ~ 2.2 eV. The prominent absorption peak appeared at 3.7 eV. Hydrosilylation, phenylation, alkoxylation and amination redshifted the absorption onset by 0.4, 0.2, 0.5 and 0.5 eV, respectively. However, the shift of the prominent absorption peak induced by organic surface modification was relatively small (in the range from 0.1 to 0.2 eV). The optical absorption demonstrated by Wang et al. qualitatively agrees with experimental results obtained by Nakano et al. The shapes of the absorption spectra for hydrosilylated and aminated silicene were consistent with those obtained by photocurrent measurements. The experimentally obtained absorption onset of phenylated and hydrosilylated silicene was ~ 3 eV and ~ 2.7 eV, respectively. The calculated absorption onset of ~2 eV (1.8 eV) for phenylated (hydrosilylated) silicene was within the range underestimated by the DFT calculations. Therefore, it can be concluded that the changes in the optical properties after organical surface modification is actually caused by surface chemistry.

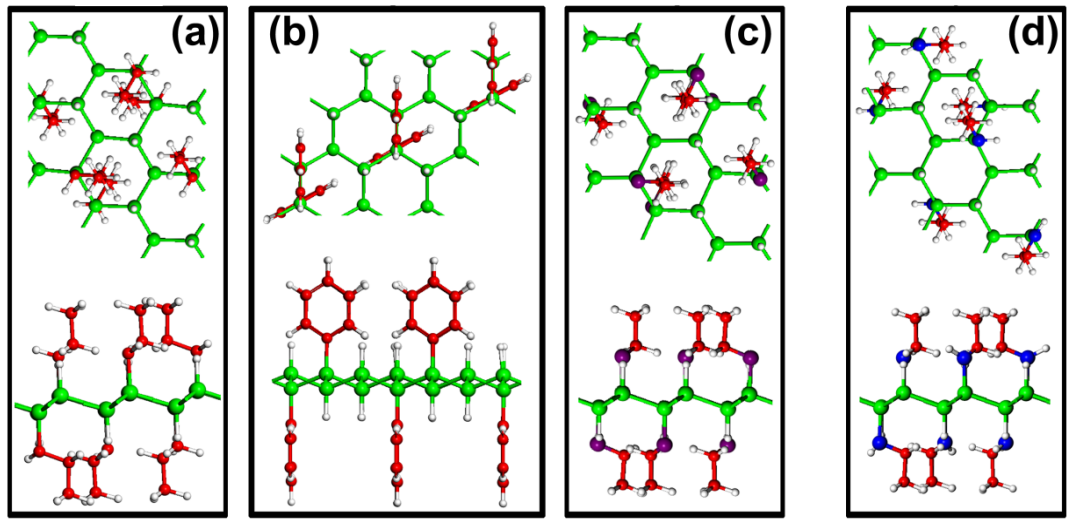

**Fig. 15** Optimized structures of (a) hydrosilylated silicene, (b) phenylated silicene, (c) alkoxylated silicene and (d) aminated silicene. Both top and side views of silicene and organically surface-modified silicene are shown. Si, H, C, O and N atoms are denoted by green, white, red, blue and purple balls, respectively [39].

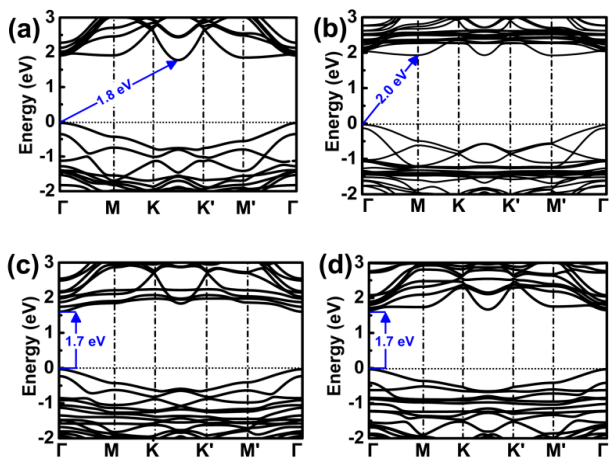

**Fig. 16** Band structures of (a) hydrosilylated silicene, (b) phenylated silicene, (c) alkoxylated silicene and (d) aminated silicene. Energy is shifted so that the Fermi level is at 0 eV [39].

Spencer et al. calculated the interaction between phenylated silicene nanosheets and the effect of the interaction on the electronic structure of phenylated silicene [45]. Firstly, they compared the adhesion energy for phenylated silicene with different separation distances using the PBE functional, the Grimme potential and the different DFT-vdWs functionals. Fig. 17 shows that there was negligible interaction between phenylated silicene nanosheets at all separation distances when the PBE functional was employed. The adhesion energy was significantly stronger as the DFT-D2 Grimme and vdW-DF2 methods were used, although the adhesion was still very weak. The optB88 gave the strongest adhesion energy at the separation distance of 10.14 Å.

The authors then calculated the band structure of phenylated silicene with different separation distances using the optB88 functional. As shown in Fig. 18 (a), phenylated silicene with infinite (27.64 Å) separation distance had a direct bandgap of 1.99 eV. This result was different from Wang et al.'s[39], in which the separation distance of phenylated silicene nanosheets was 20 Å (also could be regarded as infinite separation distance). Wang et al. used the PBE functional to calculate the band structure of phenylated silicene without considering van der Waals (vdWs) interactions. In contrast, vdWs interactions were concerned in Spencer et al.'s work with the optB88 functional. Therefore, we can see that vdWs interactions introduced remarkable changes in the electronic properties of phenylated silicene nanosheets. At the minimum interfacial separation distance (10.14 Å), phenylated silicene remained to be a direct-bandgap semiconductor with a smaller bandgap of 1.98 eV (Fig. 18 (b)). When the separation distance of phenylated silicene sheets was compressed to be 8.64 Å, phenylated silicene became an indirect-bandgap semiconductor with the bandgap of 2.01 eV (Fig. 18 (c)). This suggested that the properties of phenylated silicene could be tuned by exerting compression stress.

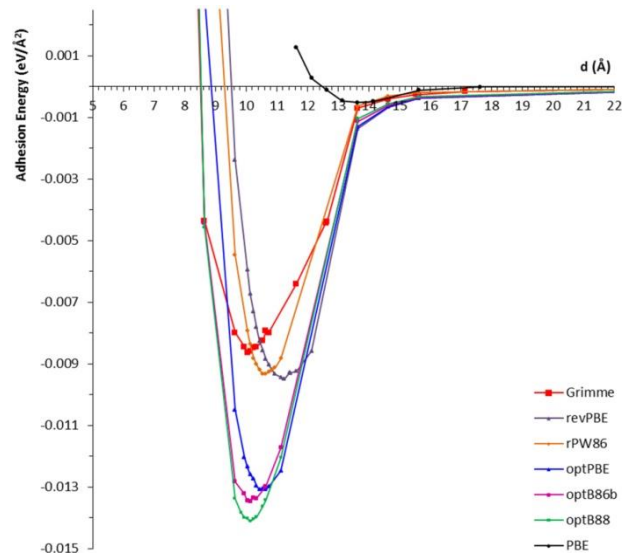

**Fig. 17** Calculated adhesion energy curves for phenylated silicene with different separation distances by. Different functionals used in the calculations are indicated in the figure [45].

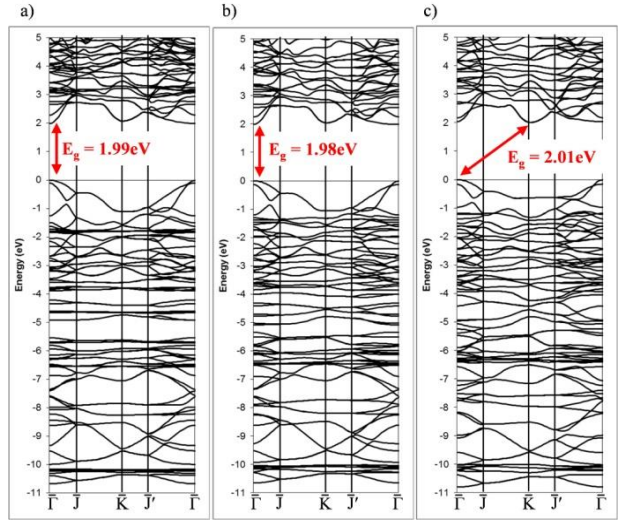

**Fig. 18** Band structures of phenylated silicene obtained by using the optB88 functional at the separation distances of (a) 27.64 Å (infinite), (b) 10.14 Å (minimum) and (c) 8.64 Å (compressed). The bandgap ($E_g$) is indicated by the red arrow [45].

## 4. Oxidation

Similar to what occurs for traditional silicon materials, the oxidation of silicene is of the great question during the fabrication of silicene-based devices. Experimental and theoretical researches have been carried out to investigate the formation of silicene oxides and the effect of oxidation on the electronic properties of silicene.

### 4.1 Experimental work

The oxidation of silicene with $O_2$ dosing in an ultrahigh vacuum (UHV) environment is different from that upon exposure to air. We now review experiments that have been carried out for these two types of silicene oxidation.

#### 4.1.1 Oxidation of silicene with $O_2$ dosing in ultra-high vacuum

Silicene exhibits strong resistance towards oxidation upon $O_2$ exposure in an UHV environment [48, 49]. Molle et al. synthesized silicene on a Ag (111) substrate by molecular beam epitaxy (MBE) [49]. Fig. 19 (a) shows that silicene has the superstructures of $\sqrt{13} \times \sqrt{13}$ and $4 \times 4$ in Molle et al.'s work. *In-situ* XPS measurements demonstrated that the Si 2*p* line related to original silicene peaked at a binding energy of 98.8 eV (Fig. 19 (b)), similar to that for bulk Si (111). After the exposure of silicene to $O_2$ at 1000 L, the shape of the Si 2*p* line exhibited a small asymmetry (Fig. 19 (c)), which could be related to a minor Si-O component. The scarce oxidation might be enabled by structural defects or domain boundaries in silicene (Fig. 19 (a)) [49].

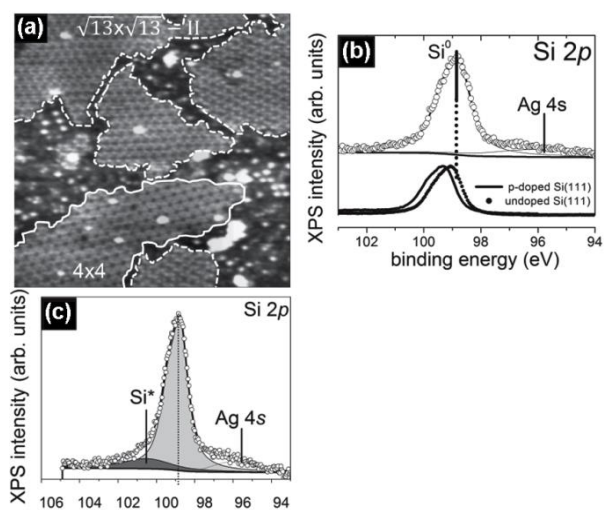

**Fig. 19** (a) 35×35 nm$^2$ STM topography of silicene on Ag (111) substrate. The STM topography illustrates the interplay of differently structured silicene domains. (b) Si 2*p* XPS spectra of silicene epitaxially grown on Ag(111). (c) Si 2*p* XPS spectra of silicene exposed to 1000 L $O_2$ [49].

Friedlein et al. investigated the oxidation of silicene on the substrate of $ZrB_2$ with $O_2$ dosing in an UHV environment [50]. They also found that the scare oxidation of silicene only started at a high dose (4500 L) of $O_2$, as shown in Fig. 20 (a). They studied the oxidation of silicene with different coverage of Al atoms. It was found that the Al coverage led to rather small changes in the Si 2*p* spectra (Fig. 20 (b)). This indicated that the interaction between Al atoms and silicene was very weak. Upon exposure to $O_2$, a broad Si 2*p* high-binding energy component locating at ~ 103 eV appeared, indicating a major fraction of Si atoms was oxidized (Fig. 20 (c)). Therefore, Al atoms could catalyze the oxidation of silicene with $O_2$ dosing in an UHV environment [50].

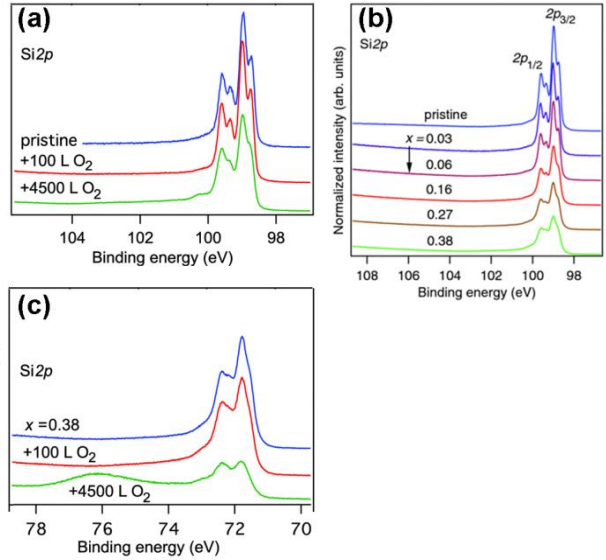

**Fig. 20** Si 2*p* XPS spectra of (a) pristine silicene before and after exposure to the $O_2$ doses of 100 L and 4500 L, (b) silicene with the surface coverage of Al atoms ranging from x = 0 (pristine) to x = 0.38 ± 0.06, (c) silicene with the surface coverage of 0.38 ± 0.06 for Al atoms before and after exposure to the $O_2$ doses of 100 L and 4500 L [50].

Fig. 21 shows Du et al.'s STM results on the oxidation of silicene. They found that an oxygen atom overbridged two adjacent Si atoms in silicene, leading to the formation of Si-O-Si bonds in silicene oxide (Fig. 21 (a)) [46]. They also studied the electronic properties of this type of so-called partially oxidized silicene by scanning tunneling spectroscopy (STS) differential conductance (dI/dV) measurements (Fig. 21 (b) and (c)). It was found that the oxidation opened the bandgap of silicene by 0.11-0.18 eV, depending on the superstructures of silicene. Moreover, the bandgap of oxidized silicene was larger at oxygen sites and became smaller in the locations away from oxygen. The bandgap was the same within the lateral distance of 3 nm around an oxygen atom, implying that oxygen could affect the electronic structure of silicene in a relatively large area [46].

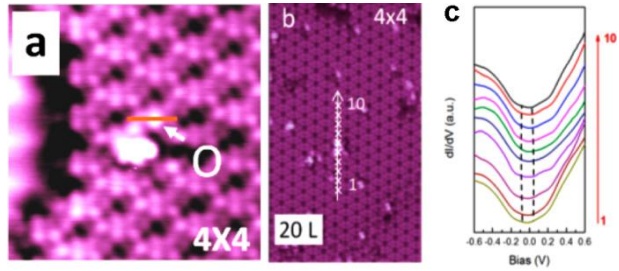

**Fig. 21** (a) STM image of partially oxidized silicene in the 4×4 structure, (b) STM image of the 4×4 silicene exposed under oxygen dose of 20 L. The arrowed line cutting across the oxygen atom represents the locations where the STS measurements are carried out, (c) the tunneling spectra (dI/dV curves) obtained along the line denoted in (b). The dashed lines in each STS result identify the bandgap [46].

Xu et al. continued Du et al.'s investigation by increasing the oxygen dose to 600 L [51]. As shown in Fig. 22 (a), the increase of the concentration of O atoms transformed the original silicene to be a disordered structure. Fig. 22 (b) shows that the Si-O peak at the binding energy of 101.6 eV clearly appeared after oxidation. The binding energy of the Si-O bond in oxidized silicene was lower than that in $SiO_2$ (102.3 eV). This suggested that the charge states of Si in the silicene oxide were lower than +4. Hence, silicene was only partially oxidized. They investigated the electronic properties of silicene before and after exposure of 600 L $O_2$ by measuring the angle resolved photoemission spectroscopy (ARPES). Fig. 22 (c) shows that there existed a weak Ag sp-band, indicating that the Ag sp-band remained stable upon the deposition of silicene. The clear hybrid surface band (HSB) located at the $k = 1.28$ Å$^{-1}$ was attributed to the hybridization of Si and Ag orbitals. The HSB indicated the metallic property of silicene. Fig. 22 (d) shows that after oxidation the metallic HSB disappeared, suggesting that the hybridization between Si and Ag was broken due to the oxidation of silicene. In the meantime, the Shockley surface state (SSS) bands resulted from the surface states of free electrons at the Ag/vacuum interface appeared. Clearly, the silicene oxide did not cover the Ag(111) substrate entirely. It was important that an asymmetric band with the highest energy at ~ -0.6 eV was observed.

This band was the valence band related to silicene oxide [51]. Therefore, it was concluded that silicene was oxidized to be a semiconductor.

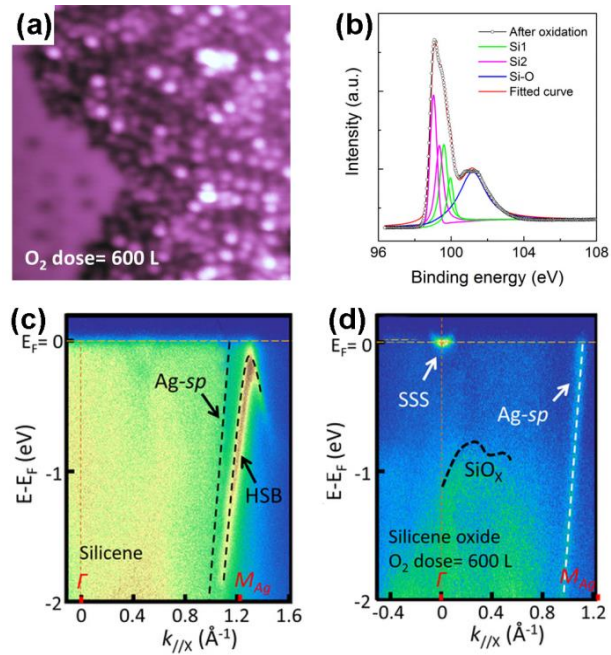

**Fig. 22** (a) STM image of the 4×4 silicene oxidized under oxygen dose of 600 L. (b) Si 2p core level XPS spectra of 4×4 silicene on Ag(111) after oxidation. Energy vs. *k* dispersion measured by ARPES for the 4×4 silicene before (c) and after (d) oxidation [51].

### 4.1.2 Oxidation of silicene upon exposure to air

The oxidation of silicene upon exposure to air is very different from that under $O_2$ exposure in ultra-high vacuum. Molle et al.'s investigation showed that the oxidation of silicene started after 3 min exposure to air [49]. The Si 2p XPS spectrum of silicene after 3 min exposure to air could be divided into three contributions, as shown in Fig. 23 (a). The peak related to the pristine silicene ($Si^0$) remained, indicating the existence of Si-Si bonds after 3 min exposure to air. The major Si-O component indicated by the binding energy of 102.3 eV was related to the $SiO_2$-like chemisorption ($Si^{4+}$ peak). The third peak ($Si^*$) was associated with the bonding of sub-oxide species with lower valence states of Si. Therefore, silicene after 3 min exposure to air was partially oxidized. However, after 1 day exposure to air, silicene was fully oxidized, because the Si 2p XPS spectrum could be totally attributed to the SiO$_2$-like chemisorption (Fig. 23 (b)).

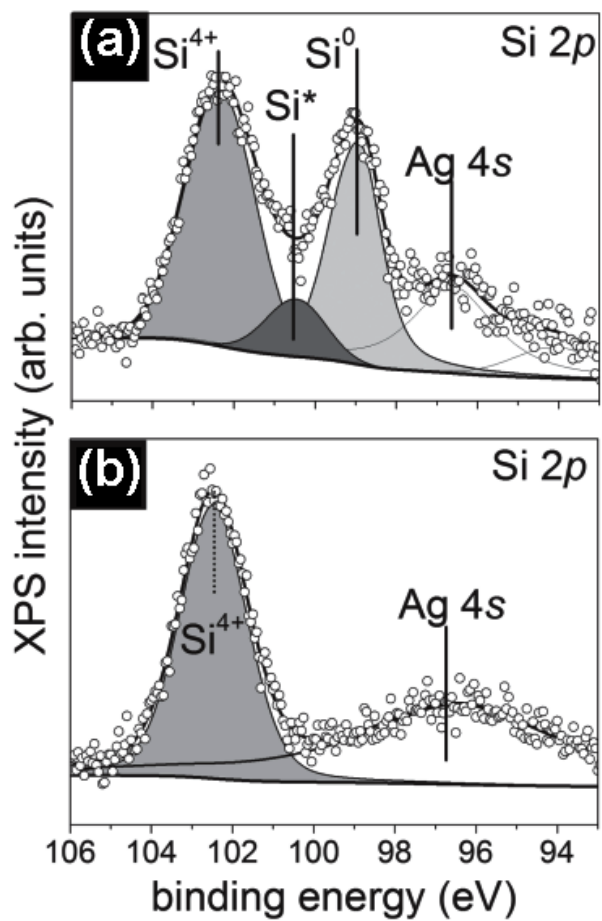

**Fig. 23** Si 2p XPS spectra of silicene (a) after 3 min exposure to air and (b) silicene after 1 day exposure to air [49].

Molle et al. further studied the deposition of Al and Al$_2$O$_3$ layers on the top of silicene for hindering the oxidation of silicene in air [49]. It was found that the Si 2p line shifted to lower binding energy after the deposition of 4 nm Al-capping layer, which was related to the extra charge exchange between the Al and Si orbitals (Fig. 24 (a)). Molle et al. investigated the oxidation of 7-nm-thick Al capped silicene upon air exposure. After oxidation, the Al 2p line had two components (Fig. 24 (b)). The Al$^*$ peak was related to partial Al oxidation at the surface, while the Al$^0$ peak was related to original Al-Al bonding at the bottom of the capping layer. The preservation of a pure Al layer at the bottom of the capping layer can prevent the underlying silicene from oxidation. Molle et al. also studied the effect of the direct deposition of $Al_2O_3$ on the oxidation of silicene. Fig. 24 (c) shows a four-component XPS line for $Al_2O_3$ capped silicene on Ag (111). The Ag 4s related peak was associated with the Ag (111) substrate. The $Si^0$ peak was related to the Si-Si bonding in silicene. The $Si^*$ peak at the binding energy of 101.6 eV was related to the partial oxidation of silicene. The $Al_2O_3$ related peak was at the binding energy of 103.3 eV. Molle et al. also investigated the oxidation of $Al_2O_3$-capped silicene upon air exposure by Raman spectroscopy. As shown in Fig. 24 (d), the narrow peak at 516 cm$^{-1}$ that was related to silicene indicated that deposition of $Al_2O_3$ layer prevented silicene from oxidation.

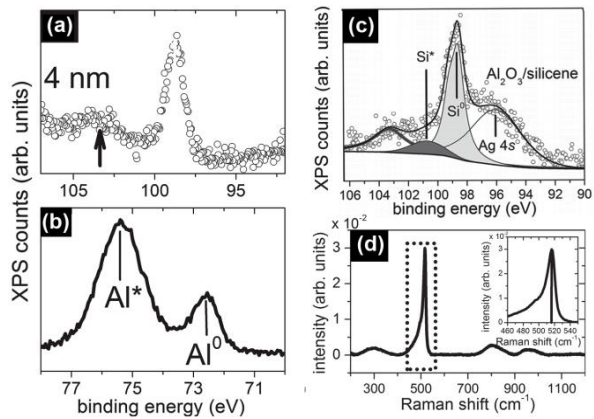

**Fig. 24** (a) Si 2p XPS line of the as-synthesized 4-nm-thick Al capped silicene, (b) Al 2p XPS line of the 7-nm-thick Al capped silicene after exposure to air, (c) Si 2p XPS line of the as-synthesized $Al_2O_3$ capped silicene, (d) Raman spectrum of the $Al_2O_3$ capped silicene exposed to air [49].

De Padova et al. showed that multilayer silicene (up to 43 layers) exhibited a small amount of surface oxidation after 1 day exposure to air. They claimed that the oxidized region at the surface served as an encapsulation layer for the rest of the silicene film [47].

**4.2 Theoretical work**

Liu et al. have studied the adsorption and dissociation of oxygen ($O_2$) molecules on silicene by first-principles calculation [29]. Their results showed that $O_2$ molecules could be easily adsorbed at the silicene surface and dissociated into O atoms without overcoming any energy barrier. The dissociated O atoms were difficult to migrate on and desorb from the silicene surface (Fig. 25). Therefore, silicene could be easily oxidized. They proposed that hydrogenation of silicene would hinder the oxidation of silicene by increasing the energy barrier of $O_2$ molecule adsorption and dissociation [29].

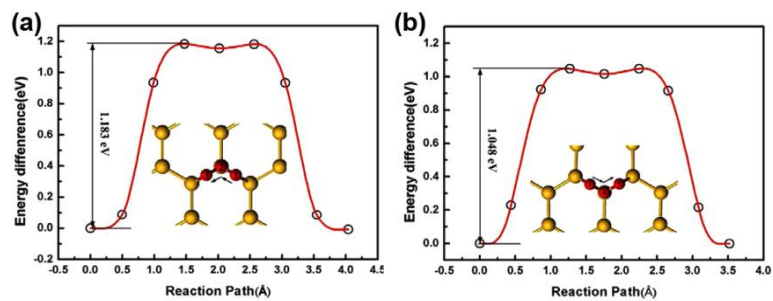

**Fig. 25** The energy difference of an O atom migrating along the pathway (a) 1-2-3 and (b) 1-4-5 on the silicene surface. (b) The insets of figures show the optimized migration pathway of 1-2-3 (a) and 1-4-5 (b) [29].

Wang et al. investigated the mechanical and electronic properties of oxidized silicene with the stoichiometric ratio of Si:O=1:1 [52]. The Poisson ratios of these silicene oxides were negative, indicating these silicene oxides exhibited unconventional auxetic behavior and thus could have obvious elastic characteristics. It was found that the silicene oxide was a direct-bandgap semiconductor with the bandgap of 0.18 eV. Meanwhile, the carrier mobility of the silicene oxide was in the order of magnitude of $10^4$ $cm^2V^{-1}s^{-1}$ [52].

Wang et al. have systematically investigated the formation, structures and electronic properties of oxidized silicene [53]. In their investigation, the oxidation of silicene was the reaction between silicene and oxygen (O) atoms or hydroxyl groups (OH). The bonding configurations of atomic O included quasi-in-plane bridging O ($O_{qb}$), double bridging O ($O_{db}$), overbridging O atoms located at either both sides ($O_{ob}^d$) or single side ($O_{ob}^s$) of the original silicene. The bonding configurations of OH included top-like OH ($OH_{tl}$), umbrella-like OH ($OH_{ul}$), boat-like OH ($OH_{bl}$) and armchair-like OH ($OH_{al}$). The charge states of Si in a silicene oxide with each above-mentioned bonding configuration ranged from +1 to +3, indicating the partial oxidation of silicene.

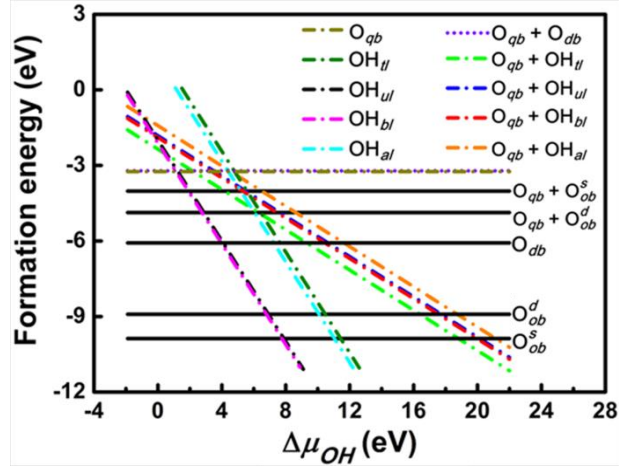

**Fig. 26** Formation energies of silicene oxides with respect to the difference between the chemical potential of OH in the oxidation system of silicene and that in water [53].

Fully oxidized silicene was obtained by incorporating $O_{qb}$ with other types of O or OH configurations. It was demonstrated that all the silicene oxides might be formed in the point of view of thermodynamics. The $E_f$ per atom of oxygen in a silicene oxide was calculated by using $E_f = (E_t - E_s - \frac{N_O}{2}\mu_{O_2} - N_{OH}\mu_{OH})/N_O$, where $E_t$ was the total energy of the SO. $E_s$ was the total energy of silicene. $N_O$ and $N_{OH}$ were the numbers of O and OH, respectively. $\mu_{O_2}$ and $\mu_{OH}$ were the chemical potentials of $O_2$ and OH, respectively. The change of $E_f$ with respect to $\Delta\mu_{OH}$ (the difference between $\mu_{OH}$ in the oxidation system of silicene and that in water) for each silicene oxide was shown in Fig. 26. For the silicene oxides induced by the partial oxidation with O atoms (e. g., in the atmosphere of oxygen gas), $E_f$ increased in the order of $O_{ob}^s$, $O_{ob}^d$, $O_{db}$ and $O_{qb}$. In De Padova et al.'s and Molle et al.'s experiments, silicene oxides with low charge states of Si were observed after the exposure of silicene to $O_2$ at 1000 L [48, [49]. Combining with Wang et al. results, it could be deduced that silicene oxides with $O_{ob}^{s}$, $O_{ob}^{d}$, $O_{db}$ were formed, the charge states of Si in these silicene oxides were +1 and +2. The $E_f$ of the silicene oxide with $O_{qb}$ is even higher than those of the silicene oxide s with $O_{qb}+O_{ob}^{s}$ and $O_{qb}+O_{ob}^{d}$. Therefore, +3 could not be the dominant charge state for Si in only O-oxidized silicene. Only O-oxidized silicene resulted in silicene oxides with the charge states of Si being +1, +2 and +4. For the silicene oxides induced by the partial oxidation with OH (e. g., in water), $E_f$ increases in the order of $OH_{bl}/OH_{ul}, OH_{al}$ and $OH_{tl}$. In Molle et al.'s experiments, the charge state of Si changed from +1 to +4 during the oxidation of silicene in air [49]. Combining with Wang et al. results, it could be concluded that that the air was an OH-rich environment. In the OH-rich environment, the value of $\Delta\mu_{OH}$ was large. When silicene is exposed to the OH-rich air, O from oxygen molecules and OH from water may both account for the oxidation. The oxidation of silicene may lead to the silicene oxides with all types of OH configurations, in which the charge state of Si is +1. $O_{qb}$ then joins the oxidation, resulting in fully oxidized silicene by means of the combination of $O_{qb}$ and one type of OH configuration. Therefore, the charge state of Si changed from +1 to +4 during the oxidation of silicene in air. These results can explain the different oxidation behaviors of silicene upon exposure to $O_2$ and air.

Wang et al. have also studied the electronic properties of oxidized silicene[51]. Their results showed that oxidation could transform silicene from semimetal to metal, semiconductor and insulator by controlling the configurations of O and OH incorporated into silicene. For semiconductive silicene oxides, the carrier motilities are less than 100 $cm^2V^{-1}s^{-1}$ [53]. The prediction of the electronic properties of silicene oxides opened an avenue of tailoring the properties of silicene through oxidation. Nevertheless, excellent control on the oxidation conditions needs to be exerted to obtain specific silicene oxides with desired electronic properties.

5.  Doping

Similar to the critical role of doping in traditional semiconductor technologies, it is believed that new technologies based on silicene may also depend on doping. For silicene, it is possible to tune the doping states by changing external fields [6-8]. Replacing Si atoms in silicene with foreign atoms (e. g., B, P, N, Al, Cr et al.) offers the possibility of tuning the structural, electronic, optical and magnetic properties of silicene. It should be noted that in this paper we define doping as replacing Si atoms in silicene with foreign atoms. Work on the adsorption of foreign atoms on silicene is not included [54-56]. In this section we will deal with the doping of silicene with both non-metallic and metallic atoms.

## 5.1 Non-metallic doping

Cheng et al. studied the effect of carrier concentration on the lattice structure, electronic structure, phonon spectrum and electro-phonon coupling of silicene by first-principles calculations [57]. They found that the lattice parameters of silicene were sensitive to the carrier concentration. As shown in Fig. 27 (a), the lattice parameter ($\alpha$) of $p$ ($n$) type silicene was smaller (larger) than that of the undoped silicene. This indicated that the lattice of silicene expanded (shrunk) when electrons were added (extracted). The buckling distance ($\Delta$) of doped silicene was larger than that of undoped silicene except for the case for $p$-type silicene with low carrier concentrations (Fig. 27 (a)).

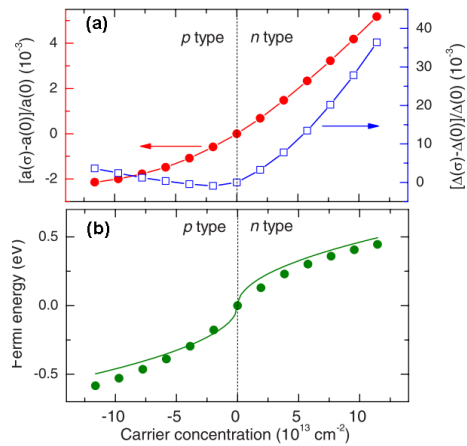

**Fig. 27** Variation of (a) the lattice parameter (σ) and buckling distance (Δ), (b) the Fermi energy as a function of the carrier concentration (σ) [57].

The lattice parameter and buckling distance were more sensitive to the carrier concentration for *n*-type doping than *p*-type doping. The Dirac cone of silicene mainly results from the $3p_z$ orbitals of Si. The bands located at 1 eV above and below the Fermi level mainly originate from the 3*s* and $3p_{x,y}$ orbitals of Si. Hence, it was predicted that doping mainly affected the occupation of the π and π* bands. Cheng et al. demonstrated that the relationship between the carrier concentration ($\sigma$) and the Fermi energy ($E_F$) is $\sigma = \text{sign}(E_F) \frac{E_F^2}{(\text{eV})^2} \times 41.48 \times 10^{13} \text{cm}^{-2}$, where $E_F = 0$ is defined by the crossing point of the π and π* bands, sign ($E_F$) is the signum of $E_F$ (Fig. 27 (b)). Phonon spectrum calculations indicated that doped silicene with the carrier concentration ranging from $-2 \times 10^{13}$ (hole-doping) to $10 \times 10^{13}$ (electron-doping) cm$^{-2}$ was stable. Moreover, the phonon dispersion indicated that Kohn anomalies at the Γ and K points were reduced by doping. Therefore, the electron-phonon coupling in doped silicene should be taken into account during its application in FETs [57].

Heavily doped silicene has also been studied. Sivek et al. have investigated the doping of silicene by using B, N and P atoms with the dopant concentration of 3.2% [58], which was achieved by replacing one Si atom in a 4 × 4 silicene supercell with a B, N or P atom. Their molecular dynamics calculations indicated that the resulting structures were all stable up to at least 500 K. Fig. 28 shows the band structures and decomposed DOS of silicene doped with B, N and P atoms obtained with DFT calculations. It was shown that the Dirac cone of silicene was preserved after the doping of B and P atoms. The Dirac cone originating from the $p_z$ orbitals of Si shifted above and below the Fermi level for B- and P- doped silicene, respectively. The shifted valance band or conduction band crossed the Fermi level for B- or P- doped silicene. Therefore, B- and P- doped silicene were metals. B and P mainly affected the deep bands of doped silicene (Fig. 28 (a) and (b)). Silicene doped with N turned out to be metallic with a band crossing the Fermi level and relative large DOS near the Fermi level. The *p* orbitals and *s* orbitals of N and Si both contributed to the bands of N-doped silicene (Fig. 28 (c)). Interestingly, N-doped silicene became ferromagnetic with a net magnetic moment of 0.9 $\mu_B$ [58].

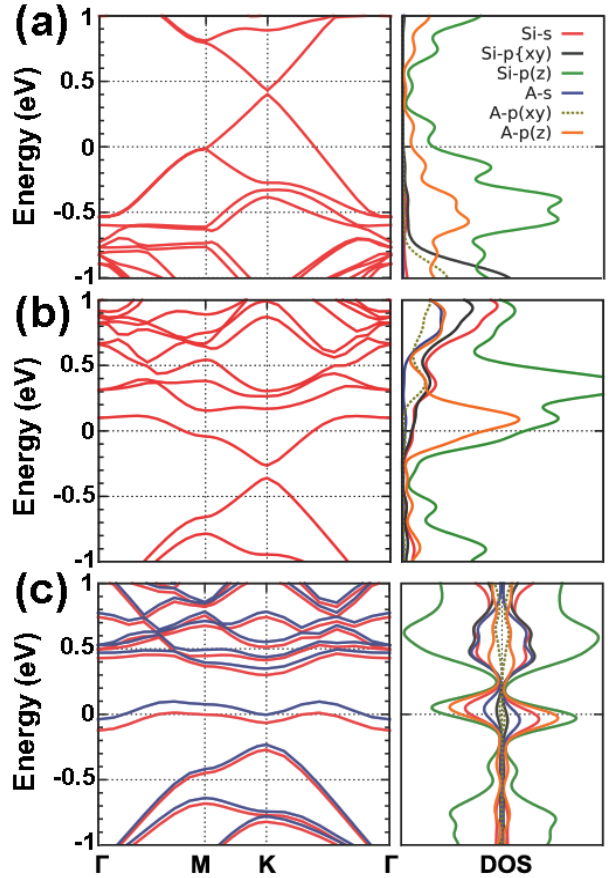

**Fig. 28** Electronic band structure and decomposed density of states (DOS) for silicene doped with (a) B, (b) P or (c) N. The Fermi level is at the energy of zero [58].

Pi et al. investigated the doping of hydrogenated silicene by using B and P with the dopant concentration ranging from 1.4% to 12.5% in the framework of DFT [59]. They found that doped hydrogenated silicene structures became direct-bandgap semiconductors with the VBM and CBM both located at the Γ point. As shown in Fig. 29, the Fermi level entered the valence (conduction) band for B (P)-doped hydrogenated silicene even when the concentration of B (P) was 1.4%, indicating that hydrogenated silicene was heavily doped even for the lowest dopant concentration of 1.4% in their work [59]. As the dopant concentration increased from 1.4% to 12.5%, the difference in energy between the Fermi level and the VBM (CBM) of B (P)-doped hydrogenated silicene became larger (i.e., 0.02 - 0.48 eV for B-doped hydrogenated silicene, 0.13-0.79 eV for P-doped hydrogenated silicene). The difference in energy between the direct bandgap and the indirect bandgap of B-doped hydrogenated silicene was smaller than that of P-doped silicene. Therefore, P doping more severely modified the band structure of hydrogenated silicene than B doping.

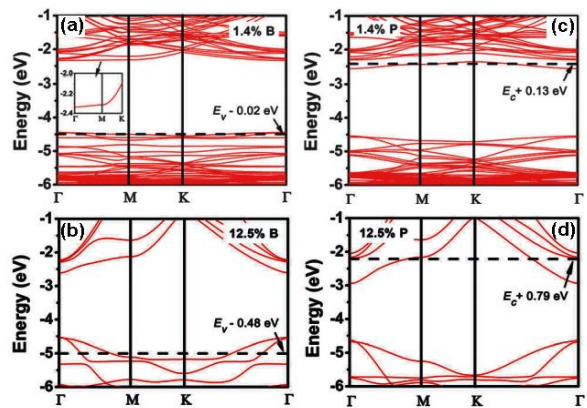

**Fig. 29** Band structures of H-silicene doped with B at the concentration of (a) 1.4%, (b) 12.5% and H-silicene doped with P at the concentration of (c) 1.4%, (d) 12.5%. The Fermi level is indicated by a horizontal dashed line [59].

Pi et al. have also calculated the effective mass ($m^*$) of holes (electrons) at the VBM (CBM) of hydrogenated silicene doped with B or P [59]. They found the largest carrier mobility could be achieved when the concentration of B was 5.6%. This might be understood by considering the smallest $m_e^*$ and $m_h^*$ for B-doped hydrogenated silicene at the B concentration of 5.6% (Fig. 30 (a)). For P-doped hydrogenated silicene, the smallest values of $m_e^*$ and $m_h^*$ could be obtained when the P concentration was 12.5%. This suggested that the largest carrier mobility for P-doped hydrogenated silicene could achieved when the P concentration was 12.5% (Fig. 30 (b)). The value of $m_h^*$ was always smaller than that of $m_e^*$ for both B- and P-doped hydrogenated silicene with the same dopant concentration, which was different from what occurs to bulk silicon. They have calculated the optical absorption spectra of hydrogenated silicene before and after P doping [59]. A subband absorption peak resulting from the electronic transition involving P-induced energy levels near the CBM occured in the low energy region. The subband absorption peak blueshifted with the increase of the dopant concentration, which was similar to what occurs to the band-edge absorption onset.

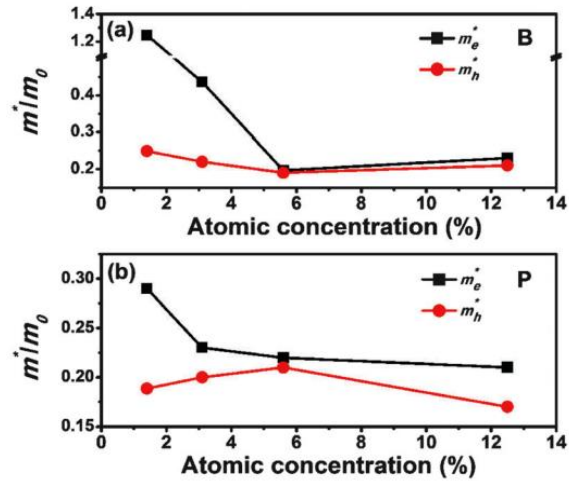

**Fig. 30** Variation of the effective mass of electrons ($m_e^*$) and holes ($m_h^*$) with the dopant concentration for H-silicene doped with (a) B or (b) P [59].

### 5.2 Metallic doping

Sivek et al. have investigated the electronic property of Al-doped silicene with the concentration of 3.2% [58]. Their molecular dynamics calculations indicated that Al-doped silicene was stable up to at least 500 K. The effect of Al doping on the electronic properties of silicene was similar to that of B doping. Al-doped silicene exhibited metallic properties. The Dirac cone derived from the $p_z$ orbitals of Si atoms shifted to be above the Fermi level. Doped Al atoms mainly affected the deep bands of Al-doped silicene [58].

Hussain et al. studied the stability, electronic structure, and hydrogen storage capacity of Mg-doped hydrogenated silicene [28]. The structure of Mg-doped hydrogenated silicene was stable in the point of view of thermodynamics. Exerting biaxial symmetric strain could increase the binding energy of Mg atoms and enhance the stability of Mg-doped H-silicene. Mg-doping changed hydrogenated silicene from a semiconductor to a metal. The electronegativity of Si is much higher than that of Mg, leading to charge transfer from Mg atoms to Si atoms. The resulting positive charge on each Mg atom can accumulate up to six $H_2$ molecules, giving rise to a $H_2$ storage capacity of 7.95% [28]. Therefore, Mg-doped hydrogenated silicene may be an ideal $H_2$ storage material.

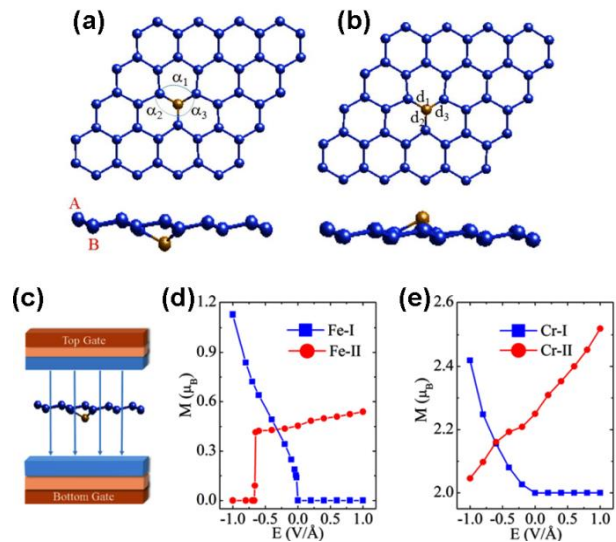

**Fig. 31** (a) The optimized structure of Fe-I in a 4 × 4 supercell with the Fe atom below the silicene plane, (b) The optimized structure of Fe-II in a 4 × 4 supercell with the Fe atom above the silicene plane, (c) The direction of the external electric field during the simulation of the magnetoelectric properties for doped silicene, (d) The magnetic moment of Fe-doped silicene for the 4 × 4 supercell as a function of the external electric field, (e) The magnetic moment of Cr-doped silicene for the 4 × 4 supercell as a function of the external electric field [57].

Zheng et al. have studied the magnetoelectric (ME) properties of silicene doped with Fe and Cr atoms by first-principles calculations [60]. They replaced one Si atom in a 4 × 4 silicene supercell with a Fe or Cr atom. The resulting dopant concentration was 3.2 %. There were two different structures with different changes of magnetism under external electric field. Silicene has a buckled structure with two sublattices denoted as the sublattices *A* and *B*. Upon doping, an Fe or Cr atom can substitute a Si atom in the sublattice *B* (*A*), resulting the X-I (X-II) structure, as shown in Fig. 31 (a) and (b). Fig. 33 (c) shows that a perpendicular field was applied to the doped silicene to investigate its magnetoelectric properties. Fig. 31 (d) illustrates that the magnetoelectric response of the Fe-I structure critically depended on the direction of the electric field. When the external electric field ($E_{ext}$) was positive, the magnetic moment (M) of the Fe-I structure was zero. When $E_{ext}$ was negative, the magnetic moment of the Fe-I structure increased linearly with the increase of the absolute value of $E_{ext}$. A clear switching effect was found in the Fe-II structure. A sharp jump of the magnetic moment from 0.41 $\mu_B$ to 0 occurred when $E_{ext}$ changed from -0.65 V/ Å to -0.66 V/Å. For the Cr-I structure, the magnetic moment was ~ 2 $\mu_B$ when $E_{ext}$ was positive. When $E_{ext}$ was negative, the magnetic moment decreased approximatively linearly with the increase of the absolute value of $E_{ext}$. For the Cr-II structure, the magnetic moment increased nearly linearly when $E_{ext}$ increased from -1 V/Å to 1 V/Å (Fig. 31 (e)). It was found that the different magnetoelectric responses of Fe- or Cr-doped silicene were related to the electronic states around the Fermi level [60]. The interesting magnetoelectric responses of Fe- or Cr-doped silicene has potential applications in magnetoelectric diodes and magnetoelectric devices with high speed response.

## 6. Formation of 2D hybrids

Formation of 2D hybrids provides an additional degree of freedom for the materials design of silicene. Since Si, C and Ge are all group IV elements, they may readily form hybrids in principle. Graphene and germanene have been synthesized. It is reasonable to expect the formation of silicene-graphene hybrids and silicene-germanene hybrids. To date, both the silicene-graphene hybrids and silicene-germanene hybrids have been theoretically investigated.

**6.1 Silicene-graphene hybrid**

Fig. 32 shows the optimized structure of a silicene-graphene (SC) hybrid with Si/C=1. Drissi et al.'s LDA-DFT calculations indicated that this type of SC hybrid was stable with a planar hexagonal lattice. The Si-C bond length was 1.78 Å. The Si-C-Si/C-Si-C bond angles (∠(Si-C-Si)/∠(C-Si-C) were 120° [61]. They found that the SC hybrid was a direct-bandgap semiconductor with the bandgap of 2.4 eV [61].

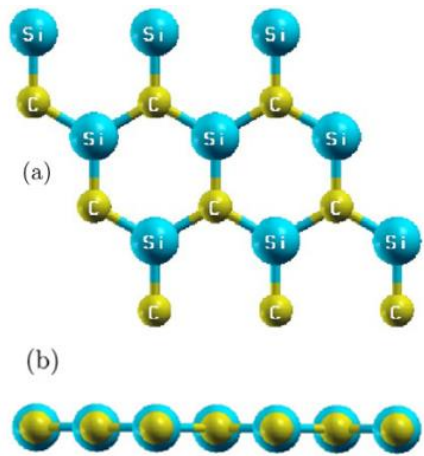

**Fig. 32** Optimized structure of the SC alloy with both (a) top and (b) side views [61].

Since DFT calculations underestimate the bandgap energies of semiconductors, Drissi et al. then calculated the band structure of the SC hybrid with the GW approximation to include the quasiparticle corrections [62]. As shown in Fig. 33 (a), the GW approximation calculation indicated the SC hybrid was a direct bandgap semiconductor. The VBM and CBM were both located at the K point. The bandgap was 3.48 eV. They also calculated the optical absorption spectra of the SC hybrid by LDA-RPA, GW-RPA (with quasiparticle corrections) and GW-BSE (with quasiparticle corrections and electron–hole interactions). The shapes of the absorption spectrum calculated by LDA-RPA and GW-RPA methods were quite similar. The GW-RPA only shifted the absorption spectra to high energies. After taking the electron-hole interactions into account, Drissi et al. found that the first absorption peak which corresponded to the optically active excitonic state emerged at 2.67 eV (Fig. 33 (b)). The exciton was mainly related to the vertical transition of electrons at the K-point between the $\pi$ state in the VBM and the $\pi^*$ state in the CBM [62].

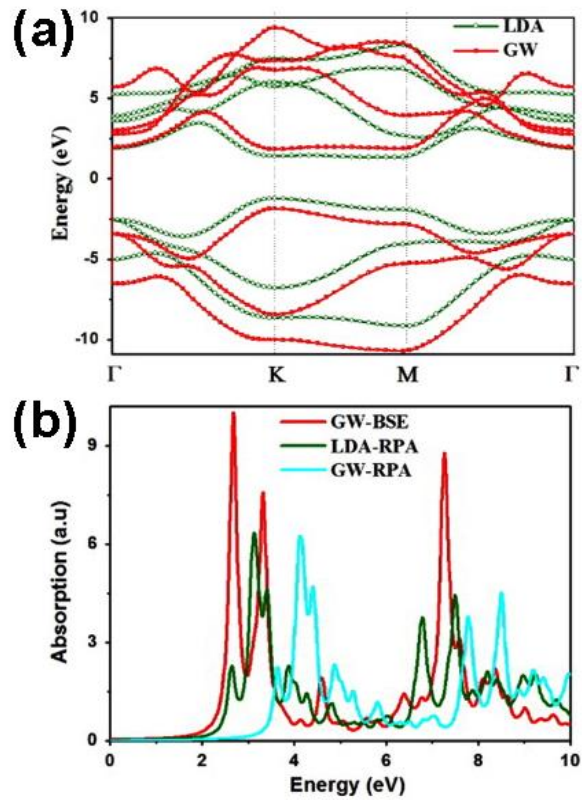

**Fig. 33** (a) Band structures and (b) optical absorption spectrum of the silicene-graphene hybrid calculated by different calculating methods [62].

Zhang et al.'s calculations revealed that the SC hybrid exhibited an excellent catalytic activity for oxygen reduction reaction without CO poisoning [63]. The SC hybrid in alkaline media had better catalytic activity than conventional Pt(111) surface. Therefore, the SC hybrid is likely used as a metal-free catalyst for oxygen reduction reaction in fuel cells.

Ding et al. investigated the geometrical and electric properties of the SC hybrid with the stoichiometric ratio of Si:C=3:1, which was shown in Fig. 34 (a) [64]. They found that the $CSi_3$ hybrid was also stable in a planar hexagonal lattice with the Si-C (Si-Si) bond length of 1.81 Å (2.26 Å). Fig. 34 (b) shows the band structure of the $CSi_3$ hybrid. It is clear that the $CSi_3$ hybrid is semi-metallic. Its CBM mainly originates from the C $p_z$ orbital, while its VBM is mainly related to the Si $p_z$ orbital [64]. This interesting phenomenon provides a peculiar example, which shows that orbitals from different elements can also form linear $\pi$ and $\pi^*$ bands around the Dirac cone.

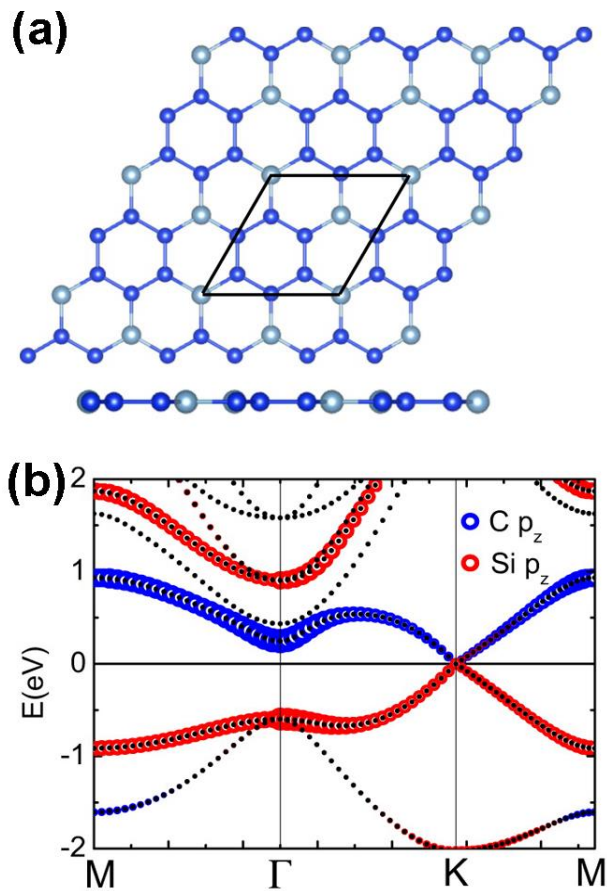

**Fig. 34** (a) Optimized structures of the $CSi_3$ hybrid with both top and side views. (b) Orbital-resolved band structure of the $CSi_3$ hybrid [64].

**6.2 Silicene-germanene hybrid**

The optimized structure of silicene-germanene (SG) hybrid is quite similar to that of the SC hybrid. However, the SG hybrid has a buckled structure, with the Si-Ge bond length of 2.38 Å and the buckling distance of 0.60 Å (Fig. 35 (a)). The band structure of the SG hybrid indicates that it is a semimetal with the top of the valence band and the bottom of the conduction band linearly crossing at the Fermi level (Fig. 35 (b)) [65].

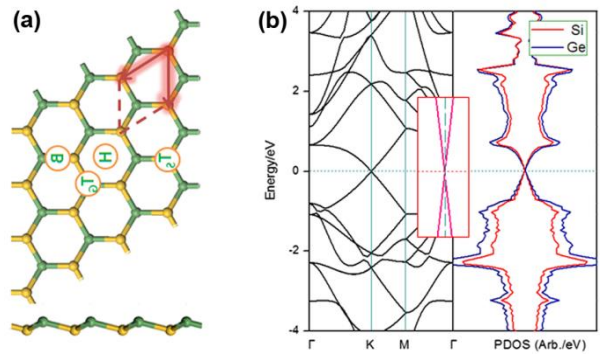

**Fig. 35** (a) Optimized structures of the silicene-germanene hybrid with both top and side views. Si and Ge atoms are denoted by yellow and green balls, respectively. (b) The band structure (left) and the partial DOS (right) of the silicene-germanene hybrid [65].

Zhang et al. have also investigated the properties of brominated SG hybrids [65]. Bromine atoms can adsorb on or passivate the SG hybrid at the sites of $T_S$, $T_G$, B and H (Fig. 35 (a)). Fig. 36 shows the resulting structures of brominated SG hybrids, which are labeled as $T_S$, $T_G$, B and H. Structure P in which Br atoms passivate the nearest neighbored Si and Ge atoms is also considered (Fig. 36). By comparing the binding energies of brominated SG hybrids with different configurations, Zhang et al. [65] found that the $T_S$ structure was the most stable. The bond lengths of Si–Ge and Si–Br in the brominated SG hybrid with the $T_S$ configuration were 2.43 Å and 2.02 Å, respectively. For the brominated SG hybrid with the $T_S$ structure the spin-up channel of had a bandgap of 1.73 eV, while the spin-down channel was characteristic of metallic bands. Upon bromination, Br atoms and Si atoms formed strong $\sigma$ bonds, leaving Ge atoms being unsaturated. The unpaired $p_z$ electrons of unsaturated Ge atoms resulted in spin polarization, which caused the net magnetic moment of 1.0 $\mu_B$. From the spin-density distributions of the brominated SG hybrid with the $T_S$ structure it was inferred that the total magnetic moment mainly resulted from unpaired Ge atoms.

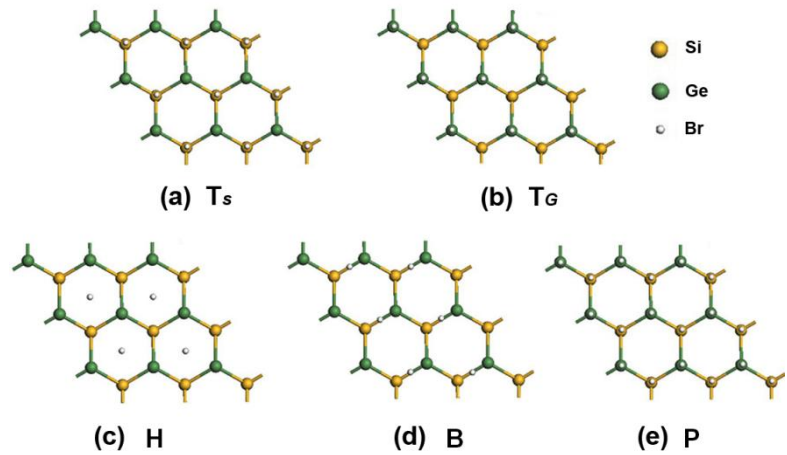

**Fig. 36** Optimized structures of brominated SG hybrids with (a) $T_S$, (b) $T_G$, (c) H, (d) B, and (e) P structures. Si, Ge and Br atoms are denoted by yellow, green and white balls, respectively [65].

## 7. Conclusion and outlook

The chemical modification of silicene is an important issue in the development of silicene. Theoretical researchers have done a lot of work on the geometrical, electronic, optical and magnetic properties of chemically modified silicene in the past few years. The successful synthesis of silicene has accelerated the experimental work on the chemical modification of silicene [79-84]. In the present work, we have reviewed the effect of hydrogenation, halogenation, organic surface modification, oxidation, doping and formation of 2D hybrids on the various properties of silicene. It has been theoretically demonstrated that chemical modification offers the possibility of tuning the properties of silicene by passivating or substituting Si atoms in silicene. Metals, semimetals, semiconductors and insulators may be produced from the single material of silicene only by means of chemical modification. However, experimental work on the chemical modification of silicene remains rather limited. There are only few experiments on the organic surface modification, oxidation and hydrogenation of silicene.

In order to facilitate the experiments on the chemical modification of silicene, further theoretical investigation needs to be carried out to search for novel properties enabled by the chemical modification of silicene and guide the design of rational routes for the chemical modification of silicene. Moreover, the interaction between chemically modified silicene layers deserves careful investigation. The stacking of chemically modified silicene may lead to novel properties. One of the most important goals of chemical modification is the opening and tuning of the bandgap of silicene and thus facilitating the application of chemically modified silicene in various device structures. In this context, simulation on silicene-based devices is highly desired. It is clear that the modulating of the properties of silicene by chemical modification depends on the accurate control of the bonding configurations. Excellent control on the chemical modification processes must be exerted to obtain specific chemically modified silicene with desired properties.

## References

[1] Kara A, Enriquez H, Seitsonen A P, Lew Yan Voon L C, Vizzini S, Aufray B and Oughaddou H 2012 *Surf. Sci. Rep.* **67** 1.

[2] Xu M, Liang T, Shi M and Chen H 2013 *Chem. Rev.* **113** 3766.

[3] Chen L, Liu C C, Feng B, He X, Cheng P, Ding Z, Meng S, Yao Y and Wu K 2012 *Phys. Rev. Lett.* **109** 056804.

[4] Feng B, Li H, Liu C C, Shao T N, Cheng P, Yao Y, Meng S, Chen L and Wu K 2013 *Acs Nano* **7** 9049.

[5] Liu C C, Feng W and Yao Y 2011 *Phys. Rev. Lett.* **107** 076802.

[6] Guzmán-Verri G and Lew Yan Voon L 2007 *Phys. Rev. B* **76** 075131.

[7] Ni Z, Liu Q, Tang K, Zheng J, Zhou J, Qin R, Gao Z, Yu D and Lu J 2012 *Nano Lett.* **12** 113.

[8] Drummond N D, Zólyomi V and Fal'ko V I 2012 *Phys. Rev. B* **85** 075423.

[9] Koski K J and Cui Y 2013 *Acs Nano* **7** 3739.

[10] Lay G L 2015 *Nat. Nano.* **10** 202.

[11] Tao L, Cinquanta E, Chiappe D, Grazianetti C, Fanciulli M, Dubey M, Molle A and Akinwande D 2015 *Nat. Nano.* **10** 227.

[12] Zandvliet H J W 2014 *Nano Today* **9** 691.

[13] Huang B, Xiang H and Wei S 2013 *Phys. Rev. Lett.* **111** 145502.

[14] Houssa M, Pourtois G, Afanas'ev V V and Stesmans A 2010 *Appl. Phys. Lett.* **97** 112106.

[15] Ding Y and Wang Y 2012 *Appl. Phys. Lett.* **100** 083102.

[16] Lew Yan Voon L C, Sandberg E, Aga R S and Farajian A A 2010 *Appl. Phys. Lett.* **97** 163114.

[17] Houssa M, Scalise E, Sankaran K, Pourtois G, Afanas'ev V V and Stesmans A 2011 *Appl. Phys. Lett.* **98** 223107.

[18] Zhang P, Li X D, Hu C H, Wu S Q and Zhu Z Z 2012 *Phys. Lett. A* **376** 1230.

[19] Wu W, Ao Z, Yang C, Li S, Wang G, Li C and Li S 2015 *J. Mater. Chem. C* **3** 2593.

[20] Wu W, Ao Z, Wang T, Li C and Li S 2014 *Phys. Chem. Chem. Phys.* **16** 16588.

[21] Yan J, Stein R, David M, Wang X and Chou M 2013 *Phys. Rev. B* **88** 121403.

[22] Zhuang J, Xu X, Du Y, Wu K, Chen L, Hao W, Wang J, Yeoh W, Wang X and Dou S 2015 *Phys. Rev. B* **91** 161409.

[23] Osborn T H, Farajian A A, Pupysheva O V, Aga R S and Lew Yan Voon L C 2011 *Chem. Phys. Lett.* **511** 101.

[24] Guzman-Verri G G and Lew Yan Voon L C 2011 *J.Phys. Conden. Matter.* **23** 145502.

[25] Wei W, Dai Y, Huang B and Jacob T 2013 *Phys. Chem. Chem. Phys.* **15** 8789.

[26] Chinnathambi K, Chakrabarti A, Banerjee A, Deb S 2012 arXiv:1205.5099.

[27] Wei W and Jacob T 2013 *Phys. Rev. B* **88** 045203.

[28] Hussain T, Chakraborty S, De Sarkar A, Johansson B and Ahuja R 2014 *Appl. Phys. Lett.* **105** 123903

[29] Liu G, Lei X L, Wu M S, Xu B and Ouyang C Y 2014 *J. Phys. Conden. Matter.* **26** 355007.

[30] Zhang R W, Zhang C W, Ji W X, Hu S J, Yan S S, Li S S, Li P, Wang P J and Liu Y S 2014 *J. Phys. Chem. C* **118** 25278.

[31] Qiu J, Fu H, Xu Y, Oreshkin A, Shao T, Li H, Meng S, Chen L and Wu K 2015 *Phys. Rev. Lett.* **114** 126101.

[32] Zhang C W and Yan S S 2012 *J. Phys. Chem. C* **116** 4163.

[33] Pan F, Quhe R, Ge Q, Zheng J, Ni Z, Wang Y, Gao Z, Wang L and Lu J 2014 *Physica E* **56** 43.

[34] Gao N, Zheng W T and Jiang Q 2012 *Phys. Chem. Chem. Phys.* **14** 257.

[35] Zhang W B, Song Z B and Dou L M 2015 *J. Mater. Chem. C* **3** 3087.

[36] Wang X, Liu H and Tu S T 2015 *RSC Adv.* **5** 6238.

[37] Ezawa M 2013 *Phys. Rev. Lett.* **110** 026603.

[38] Zheng F, Zhang C 2012 *Nanoscal Res Lett* **7** 422.

[39] Wang R, Pi X D, Ni Z, Liu Y and Yang D 2015 *RSC Adv.* **5** 33831.

[40] Okamoto H, Sugiyama Y and Nakano H 2011 *Chem. Eur. J.* **17** 9864.

[41] Nakano H, Nakano M, Nakanishi K, Tanaka D, Sugiyama Y, Ikuno T, Okamoto H and Ohta T 2012 *J. Amer. Chem. Soc.* **134** 5452.

[42] Okamoto H, Kumai Y, Sugiyama Y, Mitsuoka T, Nakanishi K, Ohta T, Nozaki H, Yamaguchi S, Shirai S and Nakano H 2010 *J. Amer. Chem. Soc.* **132** 2710.

[43] Sugiyama Y, Okamoto H, Mitsuoka T, Morikawa T, Nakanishi K, Ohta T and Nakano H 2010 *J. Amer. Chem. Soc.* **132** 5946.

[44] Rubio-Pereda P and Takeuchi N 2013 *J. Chem. Phys.* **138** 194702.

[45] Spencer M J S, Bassett M R, Morishita T, Snook I K and Nakano H 2013 *New

*J. Phys.* **15** 125018.

[46] Du Y, Zhuang J C, Liu H S, Xu X, Eilers S, Wu K H, Peng C, Zhao J J, Pi X D, See K, Peleckis G, Wang X and Dou X 2014 *ACS Nano* **8** 10019.

[47] De Padova P, Ottaviani C, Quaresima C, Olivieri B, Imperatori P, Salomon E, Angot T, Quagliano L, Romano C, Vona A, Muniz-Miranda M, Generosi A, Paci B and Le Lay G 2014 *2D Mater.* **1** 021003.

[48] De Padova P, Quaresima C, Olivieri B, Perfetti P and Le Lay G 2011 *J. Phys. D: Appl. Phys.* **44** 312001.

[49] Molle A, Grazianetti C, Chiappe D, Cinquanta E, Cianci E, Tallarida G and Fanciulli M 2013 *Adv. Func. Mater.* **24** 5088.

[50] Friedlein R, Van Bui H, Wiggers F B, Yamada-Takamura Y, Kovalgin A Y and de Jong M P 2014 *J. Chem. Phys.* **140** 204705.

[51] Xu X, Zhuang J, Du Y, Feng H, Zhang N, Liu C, Lei T, Wang J, Spencer M, Morishita T, Wang X and Dou S X 2014 *Sci. Rep.* **4** 7543.

[52] Wang Y and Ding Y 2013 *Phys. Status. Solid-R* **7** 410.

[53] Wang R, Pi X D, Ni Z, Liu Y, Lin S, Xu M and Yang D 2013 *Sci. Rep.* **3** 3507.

[54] Lin X and Ni J 2012 *Phys. Rev. B* **86** 075440.

[55] Quhe R, Fei R, Liu Q, Zheng J, Li H, Xu C, Ni Z, Wang Y, Yu D, Gao Z and Lu J 2012 *Sci. Rep.* **2** 853.

[56] Ni Z, Zhong H, Jiang X, Quhe R, Luo G, Wang Y, Ye M, Yang J, Shi J, Lu J 2014 *Nanoscale* **6** 7609.

[57] Cheng Y C, Zhu Z Y and Schwingenschlögl U 2011 *Europhys. Lett.* **95** 17005.

[58] Sivek J, Sahin H, Partoens B and Peeters F M 2013 *Phys. Rev. B* **87** 085444.

[59] Pi X D, Ni Z Y, Liu Y, Ruan Z, Xu M and Yang D 2015 *Phys. Chem. Chem. Phys.* **17** 4146.

[60] Zheng R, Lin X and Ni J 2014 *Appl. Phys. Lett.* **105** 092410.

[61] Drissi L B, Saidi E H, Bousmina M and Fassi-Fehri O 2012 *J. Phys. Conden. Matter* **24** 485502.

[62] Drissi L B and Ramadan F Z 2015 *Physica E* **68** 38.

[63] Zhang P, Xiao B B, Hou X L, Zhu Y F and Jiang Q 2014 *Sci. Rep.* **4** 3821.

[64] Ding Y and Wang Y 2013 *J. Phys. Chem. C* **117** 18266.

[65] Zhang R W, Zhang C W, Li S S, Ji W X, Wang P J, Li F, Li P, Ren M J and Yuan M 2014 *Solid State Commun.* **191** 49.

[66] Elias D C, Nair R R, Mohiuddin T M G, Morozov S V, Blake P, Halsall M P, Ferrari A C, Boukhvalov D W, Katsnelson M I, Geim A K and Novoselov K S 2009 *Science* **323** 610.

[67] Sofo J O, Chaudhari A S and Barber G D, 2007 *Phys. Rev. B: Condens. Matter Mater. Phys.* **75** 153401.

[68] Zhou J, Wang Q, Sun Q, Chen X S, Kawazoe Y and Jena P, 2009 *Nano Lett.* **9** 3867.

[69] Cahangirov S, Topsakal M, Aktürk E, Şahin H and Ciraci S 2009 *Phys. Rev. Lett.* **102**, 236804.

[70] Williamson A J, Grossman J C, Hood R Q, Puzder A and Galli G 2002 *Phys. Rev. Lett.* **89** 196803.

[71] Schwierz F 2010 *Nat. Nanotech.* **5** 487.

[72] Zhu W, Sridhar S, Liu L, Hernandez E, Donnelly V M and Economou D J 2014 *J. Appl. Phys.* **115** 203303.

[73] Nakamura Y, Mera Y and Maeda K 2002 *Surf. Sci.* **497** 166.

[74] Fukata N, Jevasuwan W, Ikemoto Y and Moriwaki 2015 *Nanoscale* **7** 7246.

[75] Kim J, Kwon M, Logeeswaran V, Grego S and Islam M 2012 *IEEE T. Nanotechnol.* **11** 782.

[76] Ma Y, Chen X, Pi X D, Yang D 2011 *J. Phys. Chem. C* **115** 12822.

[77] Ma Y, Pi X D, Yang D 2012 *J. Phys. Chem. C* **116** 5401.

[78] Nakano H, Mitsuoka T, Harada M, Horibuchi K, Nozaki H, Takahashi N, Nonaka T, Seno Y and Nakamura H 2006 *Angew. Chem. Inter. Edi.* **45** 6303.

[79] Aufray B, Kara A, Vizzini S b, Oughaddou H, Léandri C, Ealet B and Le Lay G 2010 *Appl. Phys. Lett.* **96** 183102.

[80] Lalmi B, Oughaddou H, Enriquez H, Kara A, Vizzini S b, Ealet B N and Aufray B 2010 *Appl. Phys. Lett.* **97** 223109.

[81] Feng B, Ding Z, Meng S, Yao Y, He X, Cheng P, Chen L and Wu K 2012 *Nano Lett.* **12** 3507.

[82] Fleurence A, Friedlein R, Ozaki T, Kawai H, Wang Y and Yamada-Takamura Y 2012 *Phys. Rev. Lett.* **108** 245501.

[83] Meng L, Wang Y, Zhang L, Du S, Wu R, Li L, Zhang Y, Li G, Zhou H, Hofer W A and Gao H J 2013 *Nano Lett.* **13** 685.

[84] Chiappe D, Scalise E, Cinquanta E, Grazianetti C, van den Broek B, Fanciulli M, Houssa M and Molle A 2014 *Adv. Mater.* **26** 2096.